\documentclass[12pt]{article}
\usepackage{amssymb}
\usepackage{latexsym}
\usepackage{amsmath}
\usepackage{graphicx}
\usepackage{listings}
\usepackage{cite}
\usepackage{authblk}

\usepackage{esdiff}
\usepackage{mathtools}
\usepackage{xcolor}
\def\De{\Delta}
\def\th{\theta}
\def\vp{\varphi}

\def\la{\lambda}
\def\ka{\kappa}
\def\om{\omega}
\def\al{\alpha}
\def\be{\beta}
\def\si{\sigma}

\title{\Large{Separability of Dirac equation in the STU black hole space-time: pairwise-equal charge case study}}
\author[1,2,3]{\small {M. Cveti\v{c}\thanks{e-mail: cvetic@physics.upenn.edu}}}  
\author[1,4]{\small {M. M. Stetsko\thanks{e-mail: mstetsko@gmail.com, mstetsko@upenn.edu}}}
\affil[1]{Department of Physics and Astronomy, University of Pennsylvania, Philadelphia, PA, 19104, USA}
\affil[2]{Department of Mathematics, University of Pennsylvania, Philadelphia, PA, 19104, USA}
\affil[3]{Center for Applied Mathematics and Theoretical Physics, University of Maribor, Maribor, Slovenia}
\affil[4]{Department for Theoretical Physics, Ivan Franko National University of Lviv, Lviv, UA-79005, Ukraine}

\begin{document}
\vspace*{-2cm}
\begin{flushright}
    \texttt{UPR-1326-T}
\end{flushright}
{\let\newpage\relax\maketitle}
{\abstract{We study the separability of the Dirac equation for the four-dimensional STU black hole space-time. In particular, we analyze in detail the separability conditions in the pair-wise equal charge STU black hole space-time \cite{Cvetic_PRD96}. While in the latter case the minimally coupled Dirac equation is not separable the introduction of a specific torsion term ensures the separability. The source of the torsion is the Kalb-Ramond field, which is an integral part of String Theory, but further aspects of its properties and coupling to fermionic fields remain to be studied. To derive the torsion two different approaches are used in conformally related frames, showing that the torsion is not unique. The correspondingly modified Dirac equations in the Einstein and String frames are shown to be separable. Furthermore, the massless radial and angular wave-equations are examined, they show close similarity with corresponding equations for the standard Kerr background. A generalization of the Teukolsky equation for the pair-wise equal case is conjectured. We also briefly analyze a technically sophisticated radial equation in the massive case.}}

\section{Introduction}
Recent progress in the gravitational waves detection \cite{Abbott_PRL2016,Amendola_LRR2018} stimulated considerable interest in studies of various aspects of gravitational theories  \cite{Clifton_PRep2012,Heisenberg_PRep2019}. The very important issue is related to consistency of theoretical gravitational models with modern observational tests \cite{Amendola_LRR2018,Will_LRR14}. Some of those issues have rather long history, in particular, studies of quantum fields in General Relativity settings. Those studies are important from the point of view of stability of the black hole space-times \cite{Nollert_CQG99}, caused by different types of perturbations, including gravitational ones.

The gravitational theories considered to be alternatives to the standard General Relativity should satisfy a number of important conditions. First of all, those theories should allow a well-posed initial value formulation. The second, stress-energy tensors for matter fields should satisfy sensible energy conditions. The third, stability requirements should be imposed on black hole solutions. Finally, any time-dependent field in the corresponding gravitational background should show causal propagation. Additional requirements may also be imposed for a particular study.

It is known that within String Theory, as a preeminent candidate for consistent theory of quantum gravity, at least the first two mentioned requirements are satisfied. Moreover, the black hole solutions of String Theory provide an important testing ground to study the mesoscopic and microscopic properties of black holes in a consistent theory of quantum gravity. In particular, the so called STU black holes of the effective toroidally compactified String Theory play an important role in this program due to the explicit form of the space-time metric and other field sources. The general STU black holes are specified by the mass, angular momentum, four electric and magnetic charges and they were only relatively recently obtained \cite{Chow_CQG14,Chow_PRD14_1,Chow_PRD14_2} by employing a solution generation technique\footnote{Its BPS limit was obtained in \cite{Cv_Tseyt_PRD96,Cv_Tseyt_PLB96}.}. On the other hand, in this paper we shall focus on the explicit four charges examples of STU black holes \cite{Cvetic_NPB96_1,Cvetic_PRD96,Cvetic_NPB96,Chong_NPB05,Youm_PhysRept99} and provide detailed technical results for the pair-wise equal charges case. We should note that further specialization of the Cveti\v{c}-Youm dyonic black hole is the Kerr-Sen solution \cite{Sen_PRL92,Sen_NPB94} where two of the four charges are set to zero. The Kerr-Sen solution spurred serious interest to black holes in the framework of String Theory, particularly in Supergravity \cite{Cvetic_NPB96_1,Cvetic_PRD96,
Behrndt_NPB99,Youm_PhysRept99,Cvetic_NPB96,Chong_NPB05,
Gutowski_JHEP04_1,Gutowski_JHEP04_2,Gimon_JHEP08,Chow_CQG14,Chow_PRD14_1,Chow_PRD14_2}. 

Even though the STU black hole is not treated as a physically motivated alternative to the Kerr solution, but it can rather be considered as a useful mutiparametric generalization of the latter one, obtained within the well-defined theory. The other advantages of STU models are the facts that being Lagrangian theories they in principle allow well defined initial value formulation and matter sectors of those models satisfy the strong, weak and dominant energy conditions of General Relativity.

STU black holes became important models for investigation of implications of Supergravity and in general String Theory in gravitational sector, particularly on mesoscopic properties of black holes. On the other hand, due to the rich matter sector of STU models the issue of considerable importance is the study of material fields on STU black holes backgrounds. Being multiparametric deformations of the standard Kerr solution STU black holes allow to obtain more general features of classical and quantum fields on generalized rotating black hole backgrounds, subject to consistency conditions and thus representing an important testing ground beyond the Kerr space-time.  

To the best of our knowledge only scalar fields and in particular a minimally coupled probe scalar field on rotating STU black hole space-times have been studied extensively so far. One of the most important issues in studies of any type of fields is a separability of corresponding field equations \cite{Cvetic_JHEP12_1,Cvetic_JHEP12} for suitably chosen coordinates. The separability is not just a mere technical issue making corresponding equations tractable, but it is directly related to the existence of Killing and Killing-Yano tensors, which reflect symmetry properties of the background geometry and are related to conserved charges and in quantum case. Moreover they are crucial for a general notion of integrability of various field equations \cite{Carter_PRD79,Frolov_PRL07,Frolov_LivRevRel17}.

Symmetry properties were also studied in context of establishing relations with conformal field theory (CFT), particularly near horizon geometry of the rotating STU black holes was examined \cite{Cvetic_JHEP12_1,Cvetic_JHEP12}. Studies of quantum fields in near horizon limit not only revealed ``hidden conformal symmetry'' \cite{Cvetic_JHEP12_1,Cvetic_JHEP12,Cvetic_JHEP09}, but also deepened understanding of the space-time geometry and its thermodynamic properties \cite{Cvetic_PRD97,Cvetic_NPB97,Cvetic_JHEP09}. Conformal symmetry played important role for studying stability \cite{Cvetic_PRD14}, Love numbers \cite{Cvetic_PRD22_1}, even though stability was considered in a more general setup \cite{Cvetic_PRL20,Cvetic_PRD22}. Several other issues, for instance vacuum polarization on STU black hole backgrounds were also studied \cite{Cvetic_JHEP15,Cvetic_PRD15,Cvetic_PRD17}. 
 
In contrast to the scalar sector in STU Supergravity the fermionic one is much less studied. The fermionic sector is also important to understand how fermions can be coupled to the fields in the bosonic sector. In this work our purpose is much more modest, namely we are to examine the Dirac equation on the rotating STU black hole background \cite{Cvetic_PRD96,Chong_NPB05} which is characterized by its mass, angular momentum and four independent charges, but of technical reasons we study the case when the charges are equal in pairs (pair-wise equal charges geometry). 

It should be emphasized, that nonetheless the fact that the STU solutions are known for more than twenty years, the fermionic  and other higher spin field equations were hardly studied for the STU black hole backgrounds. As far as we know only the Dirac equation in the Kerr-Sen space-time with equal electric charges was considered \cite{Houri_JHEP10} together with a related concept of torsional Killing-Yano tensor, the symmetries of the Dirac operator were studied in an accompanying paper \cite{Houri_CQG10}. We point out that similar concepts for the five dimensional String Theory inspired models appeared a little bit earlier \cite{Wu_PRD09,Wu_PRD09_2} and torsional generalization of Killling-Yano tensor was proposed \cite{Kubiznak_PLB09}. 

The Dirac equation on the Kerr-Sen space-time was shown to have additional coupling to the scalar sector of the STU model, since the torsion term which was taken into account to provide separability was associated with Kalb-Ramond field or using ``dual" descrition to the axion field \cite{Houri_CQG10} and the latter ones are essential part of String Theory. We also point out that for five dimensional case the torsion was shown to be related to Chern-Simons terms \cite{Wu_PRD09,Wu_PRD09_2}, but it can also be associated with the Kalb-Ramond field. Being an integral part of String Theory the Kalb-Ramond field gives natural contribution to the Dirac equation and this coupling should be a characteristic feature for certain fermionic fields in String Theory. In order to have more comprehensive picture of this coupling, its peculiar features the fermionic sector of the corresponding Supergravity models should be studied. But there is one of the most important consequences caused by this additional coupling which can already be claimed, namely the torsion gives rise to the generalized Killing-Yano tensors and it modifies symmetry properties of the Dirac equation \cite{Houri_CQG10,Frolov_LivRevRel17}.

We consider the Dirac equation on the rotating STU black hole background, for the so-called pair-wise equal charges case, when the electric and magnetic charges are set equal in pairs. Due to lack of studies in this vein, it can be considered as a further step in examination of fermionic fields on STU solutions backgrounds in comparison with the paper \cite{Houri_JHEP10}, where a particular case of the Kerr-Sen solution was studied. At the same time deriving more general results than in \cite{Houri_JHEP10} the current work gives more general and at the same time more detailed view of the transformational properties of the Dirac equation and corresponding wave-functions due to transformation of the frames we work with. 

This paper is organized as follows. In the next section we give a sketchy analysis of the black hole metric in the Einstein frame. In the third section the massive Dirac equation in the Einstein frame is considered, namely after a modification its separability is demonstrated. The Dirac equation in the String frame is examined in the forth section, some peculiarities related to the frame transformation are also discussed. In the fifth section we study the equations for the radial and angular components of the Dirac wave-function in the massless case. In the sixth section a brief analysis of the radial wave equations for the massive case is given. Finally, the last section contains some conclusions and discussion of obtained results. There are also two appendices which clarify some technical for the current study, but on the other hand conceptual issues.

\section{The four dimensional rotating STU black hole in the Einstein frame}

Four-dimensional STU Supergravity can be derived in the heterotic formulation, namely ten-dimensional Supergravity is toroidally reduced on $T^6$ with appropriate truncations. It allows to derive four-dimensional ${\cal N}=2$ Supergravity which is coupled to three vector multiplets. The other approach which gives rise to the four-dimensional STU Supergravity is based on reduction of eleven-dimensional Supergravity on $S^7$ together with appropriate truncation from ${\cal N}=8$ to ${\cal N}=2$ supersymmetry and turning off the gauge coupling constant. We point out that within classical Supergravity STU solutions have $SL(2,\mathbb{R})^3$ symmetry and any of those $SL(2,\mathbb{R})$ corresponds to a duality, namely dilaton-axion, complex Kahler form and complex structure field. The $SL(2,\mathbb{R})^3$ might be broken down to $SL(2,\mathbb{Z})^3$ by quantum corrections or quantization of charges.

The bosonic sector of the ${\cal N}=2$ Supergravity coupled to three vector multiplets in four-dimensional case can be casted in the form \cite{Chong_NPB05,Chow_PRD14_1,Chow_PRD14_2}:
\begin{multline}\label{lagrang_gen}
{\cal L}_{4}=R*1-\frac{1}{2}\sum^{3}_{j=1}\left(*d\phi_{j}\wedge d\phi_{j}+e^{2\phi_{j}}*d\chi_{j}\wedge d\chi_{j}\right)-\frac{1}{2}e^{-\phi_1}\left(e^{\phi_2-\phi_3}*F_{1}\wedge F_{1}+\right.\\\left.e^{\phi_2+\phi_3}*F_2\wedge F_2+e^{\phi_3-\phi_2}*{\cal F}_{1}\wedge{\cal F}_{1}+e^{-\phi_2-\phi_3}*{\cal F}_{2}\wedge{\cal F}_{2}\right)-\chi_{1}\left(F_{1}\wedge{\cal F}_{1}+F_{2}\wedge{\cal F}_{2}\right).
\end{multline}
 where $R$ is the scalar curvature, $\phi_{j}$ and $\chi_{j}$ denote dilaton and axion fields respectively, and the $F_{i}$ and ${\cal F}_{i}$,  $i=1,2$ are the gauge field strengths defined in terms of potentials, namely they can be written as follows:
 \begin{eqnarray}
 F_1 &=& dA_{1}-\chi_{2}d{\cal A}_2;\\ F_{2} &=& dA_{2}+\chi_{2}d{\cal A}_{1}-\chi_{3}dA_{1}+\chi_{2}\chi_{3}d{\cal A}_{2};\\{\cal F}_{1} &=& d{\cal A}_{1}+\chi_{3}d{\cal A}_{2}; \\{\cal F}_{2} &=& d{\cal A}_{2}.
 \end{eqnarray}
We point out that other forms of the model, its symmetries and dualities are given in \cite{Chow_PRD14_2}. It is known that the four-dimensional Lagrangian (\ref{lagrang_gen}) can be obtained via reduction of a six dimensional Supergravity Lagrangian which includes  the Ricci scalar, dilaton and Kalb-Ramond fields, more details about the reduction and relations to other approaches can be found in \cite{Cremmer_NPB98,Cvetic_JHEP21,Chong_NPB05,Chow_PRD14_2}.

The model defined by the Lagrangian (\ref{lagrang_gen}) allowed to construct  a rotating black hole solution with two electric and two magnetic charges  \cite{Cvetic_NPB96_1,Cvetic_PRD96,Chong_NPB05}. Recently its generalization was also used to derive a more general solution with eight $U(1)$ charges (four electric and four magnetic charges) \cite{Chow_PRD14_1,Chow_PRD14_2}. To derive the solution with multiple charges the Kerr space-time is used as a seed solution, therefore the obtained space-time can be treated as a natural generalization of the Kerr one, as well as the generalization of earlier derived Kerr-Sen solution where only electric charges are taken into account \cite{Sen_PRL92,Sen_NPB94}.

The rotating black hole space-time with four independent $U(1)$ charges \cite{Cvetic_PRD96} can be written as a $4D$ fibration over $3D$ base space \cite{Chong_NPB05,Cvetic_JHEP12} and in the Einstein frame it takes the form:
\begin{equation}\label{metric}
ds^4=-\De^{-1/2}_{0}G\left(dt+{\cal{A}}\right)^2+\De^{1/2}_{0}\left(\frac{dr^2}{X}+d\th^2+\frac{X}{G}\sin^{2}\th d\vp^2\right),
\end{equation}
where 
\begin{eqnarray}
X &=& r^2-2mr+a^2,\label{X}\\ G &=& r^2-2mr+a^2\cos^2{\th},\\ {\cal{A}} &=& \frac{2ma\sin^2{\th}}{G}\left((\Pi_{c}-\Pi_{s})r+2m\Pi_{s}\right)d\vp,
\end{eqnarray}
\begin{multline}
\De_{0}=\prod^{3}_{I=0}\left(r+2m\sinh^2\delta_I\right)+2a^2\cos^2{\th}\left(r^2+mr\sum^{3}_{I=0}\sinh^2\delta_{I}+\right.\\\left. 4m^2(\Pi_{c}-\Pi_{s})\Pi_{s}-2m^2\sum_{I<J<K}\sinh^2{\delta_{I}}\sinh^2{\delta_{J}}\sinh^2{\delta_{K}}\right)+a^4\cos^4{\th},\label{factor}
\end{multline}
where the following abbreviations are used:
\begin{equation}
\Pi_{c}=\prod^{3}_{I=0}\cosh{\delta_I}, \quad \Pi_{s}=\prod^{3}_{I=0}\sinh{\delta_I},
\end{equation}
and here $\delta_I$, $I=0,1,2,3$ denote the $U(1)$ charges. We note, that explicit relations for all the bosonic sector fields of the model (\ref{lagrang_gen}) are given in \cite{Chong_NPB05}. The asymptotic charges of the black hole can be parametrized as follows:
\begin{eqnarray}
\nonumber G_{4}M=\frac{1}{4}m\sum^{3}_{I=0}\cosh{2\delta_{I}}, \quad G_{4}J=ma\left(\Pi_{c}-\Pi_{s}\right),\\ G_{4}Q_{I}=\frac{1}{4}m\sinh{2\delta_{I}},\quad  I=0,1,2,3,
\end{eqnarray}
and here $G_4$ is a four-dimensional gravitational constant.

In the following we consider a particular case of the metric (\ref{metric}), namely when $\delta_0=\delta_2$ and $\delta_1=\delta_3$ that is the so-called pair-wise equal charges case, the general solution will be examined elsewhere.  For this particular case, the factor $\De_0$ can be reduced to considerably simpler form, namely:
\begin{equation}
\De_0=\left((r+2m\sinh^2{\delta_1})(r+2m\sinh^2{\delta_2})+a^2\cos^2{\th}\right)^2
\end{equation}
If all the charges are set equal to each other the metric (\ref{metric}) and the factor $\De_0$ reduces to the dyonic solution with equal electric and magnetic charges and if all the charges are set to zero the metric turns to be the Kerr one.

The Boyer-Lindquist form of metric for the Kerr black hole is very convenient for various applications, in particular it makes the separation of variables for wave equations more transparent and easier to handle. We also note that for the minimally coupled scalar field \cite{Cvetic_JHEP12} a particularly chosen frame for the metric (\ref{metric}) is not important, since it is sensitive to the metric only, but not to the frame, similar conclusion is valid also for the Hamilton-Jacobi equation.  The given form of the metric (\ref{metric}) is not the Boyer-Lindquist one, but after simple transformations it can be rewritten in the desirable form, namely we obtain:
\begin{multline}
ds^2=-\frac{X}{\De^{1/2}_{0}}\left(dt+\frac{1}{G}\left(\bar{A}-a\De^{1/2}_{0}\sin^2{\th}\right)d\vp\right)^2+\\\De^{1/2}_{0}\left(\frac{dr^2}{X}+d\th^2\right)+\frac{\sin^2{\th}}{\De^{1/2}_0}\left(a dt-\frac{1}{G}\left(\De^{1/2}_{0}X-a\bar{A}\right)d\vp\right)^2,\label{BL_general}
\end{multline}
where $\bar{A}=2ma\left((\Pi_{c}-\Pi_{s})r+2m\Pi_{s}\right)\sin^2{\th}$. For the particular case when the charges are equal in pairs the Boyer-Lindquist form (\ref{BL_general}) can be considerably simplified, namely we obtain:
\begin{multline}
ds^2=-\frac{X}{\De^{1/2}_{0}}\left(dt-a\sin^2{\th}d\vp\right)^2+\De^{1/2}_{0}\left(\frac{dr^2}{X}+d\th^2\right)+\\\frac{\sin^2{\th}}{\De^{1/2}_0}\left(adt-\left((r+2ms^2_1)(r+2ms^2_2)+a^2\right)d\vp\right)^2,\label{BL_pair}
\end{multline}
where $s_i=\sinh{\delta_i}, i=1,2$ for simplicity. It should be noted that here only the the factor $\De_0$ depends on both $r$ and $\th$ while in the initial form (\ref{metric}) even for this particular case, two functions, namely $G$ and $\De_0$ depend on the both arguments $r$ and $\th$. It is easy to check that if $\delta_1=\delta_2=0$ the metric (\ref{BL_pair}) reduces exactly to the Boyer-Lindquist form of the Kerr metric. As it was noted above, in the following we will examine only pair-wise equal charges case (\ref{BL_pair}), leaving the general case for further studies.

Having the metric for this case rewritten in the Boyer-Lindquist form (\ref{BL_pair}) it is easy to extract the corresponding tetrad which will be used in the following calculations:
\begin{eqnarray}\label{tetrad_0}
\nonumber e^{0}=\frac{\sqrt{X}}{\De^{1/4}_0}\left(dt-a\sin^2\th d\vp\right),\quad e^{1}=\frac{\De^{1/4}_0}{\sqrt{X}}dr,\\ e^2 =\De^{1/4}_{0}d\th, \quad e^{3}=\frac{\sin{\th}}{\De^{1/4}_0}\left(adt-F(r)d\vp\right),
\end{eqnarray}
where $F(r)=(r+2ms^2_{1})(r+2ms^2_{2})+a^2$. We point out that upper indices in the left-hand sides of the relations (\ref{tetrad_0}) enumerate the frame fields. Obviously the Boyer-Lindquist tetrad (\ref{tetrad_0}) is Lorentzian. It should be noted that Boyer-Lindquist tetrad can be written for the general metric represented in the suitable form (\ref{BL_general}). 

The pairwise equal black hole space-time can be also derived as a solution of field equations for a truncated version of the Lagrangian (\ref{lagrang_gen}) which in Einstein frame takes the form \cite{Chong_NPB05}:
\begin{multline}
{\cal L}=R*1-\frac{1}{2}*d\phi\wedge d\phi-\frac{1}{2}e^{2\phi}*d\chi\wedge d\chi-\frac{1}{2}e^{-\phi}\times\\\left(*F_1\wedge F_1+*F_2\wedge F_2\right)-\frac{1}{2}\chi\left(F_1\wedge F_1+F_2\wedge F_2\right),\label{Lagr_1}
\end{multline} 
where for simplicity we denoted $\phi\equiv\phi_1$ and $\chi\equiv\chi_1$ and the other dilaton and axion fields are set equal to zero, we also note that the latter condition gives rise to the relation for gauge fields $F_i={\cal{F}}_i$, $i=1,2$ since corresponding charges become equal in this limit. We also note that the truncation giving rise to the Lagrangian (\ref{Lagr_1}) is consistent \cite{Chong_NPB05}. 

Using a Legendre transformation for the axion field $\chi$, or equivalently a dualization procedure one can rewrite the Lagrangian  (\ref{Lagr_1}) in the form, where instead of the axion field the Kalb-Ramond field appears. The latter form of the Lagrangian as we will see in the Section \ref{section_4} is more convenient for separation of variables in the Dirac equation. Details of the  transformation of the axion field  are given in the Appendix \ref{app_B}.

\section{The Dirac equation in the Einstein frame}
In a curved space-time the massive Dirac equation for a neutral particle takes the form:
\begin{equation}
\hat{\gamma}^{\mu}\left(\partial_{\mu}+\Gamma_{\mu}\right)\Psi+\mu_e\Psi=0,
\end{equation}
where $\hat{\gamma}^{\mu}$ are space-time gamma matrices, $\Gamma_{\mu}$ denotes space-time components of the spinor-connection $1$--form and $\mu_e$ is the fermion mass. Space-time gamma matrices $\hat{\gamma}^{\mu}$ can be decomposed in terms of more convenient Lorentzian gamma matrices $\hat{\gamma}^{A}$ as follows:
\begin{equation}\label{gamma_matr}
\hat{\gamma}^{\mu}=e^{\mu}_{A}\hat{\gamma}^{A},
\end{equation}
where $e^{\mu}_{A}$ are components of an inverse tetrad. The gamma matrices $\hat{\gamma}^{A}$ satisfy the standard anticommutation relation:
\begin{equation}
\left\{\hat{\gamma}^{A},\hat{\gamma}^{B}\right\}\equiv\hat{\gamma}^{A}\hat{\gamma}^{B}+\hat{\gamma}^{B}\hat{\gamma}^{A}=2\eta^{AB}
\end{equation}
and here $\eta^{AB}$ is the inverse of the Lorentzian metric $\eta_{AB}$. Here the Lorentzian metric is chosen to take mostly plus signature, namely $\eta_{AB}=diag(-1,+1,+1,+1)$. The Lorentzian gamma matrices are chosen as follows \cite{Wu_PRD08}:
\begin{eqnarray}\label{gamma_matr}
\hat{\gamma}^0=i\begin{pmatrix}
0 & \hat{I}\\
\hat{I} & 0
\end{pmatrix},\quad 
\hat{\gamma}^1=i\begin{pmatrix}
0 & \hat{\sigma}_3\\
-\hat{\sigma}_3 & 0
\end{pmatrix},\quad 
\hat{\gamma}^2=i\begin{pmatrix}
0 & \hat{\sigma}_1\\
-\hat{\sigma}_1 & 0
\end{pmatrix},\quad 
\hat{\gamma}^3=i\begin{pmatrix}
0 & \hat{\sigma}_2\\
-\hat{\sigma}_2 & 0
\end{pmatrix},
\end{eqnarray} 
where $\hat{I}$ is the $2\times 2$ identity matrix and $\sigma_i$, $i=1,2,3$ are the Pauli matrices. The spinor connection $\Gamma$ is defined in the following form:
\begin{equation}\label{sp1form}
\Gamma=\frac{1}{8}\left[\hat{\gamma}^A,\hat{\gamma}^B\right]\omega_{AB}=\frac{1}{4}\hat{\gamma}^A\hat{\gamma}^B\omega_{AB}
\end{equation}
and here $\omega_{AB}$ are components of a spin-connection $1$--form. The latter ones satisfy the torsion-free Cartan equation:
\begin{equation}\label{Cartan_eq}
de^{A}+{\omega^{A}}_{B}\wedge e^B=0,
\end{equation}
where $e^{A}=e^{A}_{\mu}dx^{\mu}$ is the tetrad (frame field) for our metric. Having the relation (\ref{sp1form}) it is easy to write the space-time components of the spinor connection, namely:
\begin{equation}\label{spinor_con}
\Gamma_{\mu}=\frac{1}{4}\hat{\gamma}^A\hat{\gamma}^B\omega_{AB\mu}.
\end{equation}
Using the given in the Appendix \ref{app_A} expressions for the spin-connection $1$-forms one can obtain the explicit expression for the contracted product $\hat{\gamma}^{\mu}\Gamma_{\mu}$:
\begin{multline}
\hat{\gamma}^{\mu}\Gamma_{\mu}=\frac{1}{2\De^{1/2}_{0}}\left(\left(\De^{1/4}_{0}\sqrt{X}\right)'\hat{\gamma}^{1}+\frac{1}{\sin{\th}}\left(\De^{1/4}_0\sin{\th}\right)_{,\th}\hat{\gamma}^{2}\right.\\\left.+\frac{aF'}{2\De^{1/4}_{0}}\sin{\th}\hat{\gamma}^{3}\hat{\gamma}^{0}\hat{\gamma}^{1}-\frac{a\sqrt{X}}{\De^{1/4}_0}\cos\th\hat{\gamma}^{3}\hat{\gamma}^{0}\hat{\gamma}^{2}\right).
\end{multline}

As we have noted above, the standard Dirac equation in the Kerr-Sen space-time was not separable \cite{Houri_JHEP10}, nonseparability was also confirmed for the five-dimensional Dirac equation for another Supergravity background \cite{Wu_PRD09}. To cure the difficulty a specific term dubbed as the torsion, due to its complete antisymmetry, was introduced  \cite{Houri_JHEP10,Kubiznak_PLB09}. We point out that the source of the torsion is the Kalb-Ramond field which is an essential part of String Theory and more persuasive confirmation of this fact will be given in the following section. Therefore the torsion form reflects additional coupling betwen the Dirac and Kalb-Ramond or axion fields. To have more comprehensive understanding of this coupling, especially in different frames a thorough study of the fermionic sector is needed, but it goes far beyond the main scope of this work.

On the other hand, separability of the Dirac equation in a curved space-time is naturally related to the existence of a Killing-Yano tensor and its conformal cousin \cite{Carter_PRD79,Benn_CQG04,Frolov_LivRevRel17}. It was shown that for String Theory backgrounds the standard Killing-Yano tensor (KYT) or conformal Killing-Yano tensor (CKYT) should be replaced by the generalized Killing-Yano tensor (GKYT) or generalized conformal Killing-Yano tensor (GCKYT) respectively \cite{Kubiznak_PLB09,Houri_JHEP10}. GCKYT can be written in terms of standard tensor notations or more concise differential forms notations, what makes its definition more transparent. Namely, for the generalized conformal Killing-Yano $k$-form $\omega$ in a $n$-dimensional space-time we write:
\begin{equation}\label{KY_eq_form}
\nabla^{T}_{{\bf X}}\omega-\frac{1}{k+1}{i}_{{\bf X}}d^{T}\omega+\frac{1}{n-k+1}{\bf X}^{\flat}\wedge\delta^{T}\omega=0,
\end{equation}
where ${\bf X}$ is a vector field, defined on the space-time manifold (or more precisely on a tangent bundle over the space-time manifold),  $i_{{\bf X}}$ is the interior product, ${\bf X}^{\flat}$ denotes a $1$-form constructed via the canonical (``musical'') isomorphism for the field ${\bf X}$, symbols $\nabla$, $d$ and $\delta$ correspond to covariant, exterior derivatives and coderivative respectively and the superscript $T$ denotes their torsion modified counterparts which are defined as follows:
\begin{align}
\nabla^{T}_{{\bf X}}\omega &=\nabla_{{\bf X}}\omega+\frac{1}{2}i_{\bf X}T\underset{1}\wedge\omega;\\ d^{T}\omega &=d\omega-T\underset{1}\wedge\omega;\\ \delta^{T}\omega &=\delta\omega-\frac{1}{2}T\underset{2}\wedge\omega;
\end{align}
and in the right hand sides of the upper relations $T$ is the torsion form and specific contracted wedge products are used. The contracted wedge product is defined as follows:
\begin{equation}\label{nf_wedge}
\chi\underset{0}\wedge\eta=\chi\wedge\eta, \quad \chi\underset{j}\wedge\eta=i_{{\bf X}^{a}}\chi\underset{j-1}\wedge i_{{\bf X}_{a}}\eta,
\end{equation} 
here we point out that vector fields ${\bf X}_{a}$ form an orthonormal basis $g({\bf X}_{a},{\bf X}_{b})=\eta_{ab}$ and there is a summation over $a$ in the contracted product (\ref{nf_wedge}).

Instead of differential form representation for the generalized Killing-Yano form the standard tensor notations can be used. In particular, since we are interested in the Killling-Yano tensors of the rank two in the four dimensional space-time, the generalized conformal Killing-Yano tensor $k_{\mu\nu}$ satisfies the following equation:
\begin{equation}\label{gkyt}
\nabla^{(T)}_{\mu}k_{\lambda\kappa}=\nabla_{[\mu}^{(T)}k_{\lambda\kappa]}+\frac{2}{3}g_{\mu[\lambda}\nabla^{(T)}_{|\sigma|}{k^{\sigma}}_{\kappa]},
\end{equation}
where $\nabla^{(T)}_{\mu}$ is the torsion-modified covariant derivative which acts on a tensor $W_{\lambda\kappa}$ as follows:
\begin{equation}\label{cov_der}
\nabla^{(T)}_{\mu}W_{\la\ka}=\nabla_{\mu}W_{\la\ka}+\frac{1}{2}{T_{\mu\la}}^{\sigma}W_{\sigma\ka}-\frac{1}{2}{T_{\mu\ka}}^{\sigma}W_{\la\sigma}.
\end{equation}
The bracket $[]$ for subscript indices in (\ref{gkyt}) means antisymmetrization over the indices enclosed by the bracket.

Using the tetrad basis (\ref{tetrad_0}) we write the generalized conformal Killing-Yano two-form $\omega$ and its Hodge dual, namely generalized Killing-Yano form $f$ in the following way:
\begin{eqnarray}\label{KY_form}
\label{om_Ef}\omega_{\pm}=\sqrt{(r+2ms_1^2)(r+2ms_2^2)}e^{0}\wedge e^{1}\pm a\cos{\theta}e^{2}\wedge e^{3};\\\label{f_Ef} f_{\pm}=-\sqrt{(r+2ms_1^2)(r+2ms_2^2)}e^{2}\wedge e^{3}\pm a\cos{\theta}e^{0}\wedge e^{1}.
\end{eqnarray}
We point out here that for the particular cases when all the charges are set equal (equal charges dyonic solution) or set to zero (Kerr case) the generalized Killing-Yano forms (\ref{om_Ef}) and (\ref{f_Ef}) are reduced to the ordinary Killing-Yano forms (or tensors in tensor notations) satisfying the standard torsionless Killing-Yano equations.

We also note that for the torsionless forms, for instance for the mentioned above particular cases the conformal Killing-Yano form $\omega_{\pm}$ is closed, namely $d\omega_{\pm}=0$ and the Killing-Yano form $f_{\pm}$ is co-closed $\delta f_{\pm}=0$, therefore it simplifies the torsionless counterpart of the equation (\ref{KY_eq_form}). Since we take into account the torsion, to maintain the structure of the Killing-Yano equation for a particularly chosen Killing-Yano form it is natural to assume that instead of the closed and co-closed forms there are a $T$-closed form $\omega_{\pm}$  and $T$-co-closed form $f_{\pm}$ respectively, namely it means that $d^{T}\omega_{\pm}=0$ and $\delta^{T}f_{\pm}=0$ correspondingly.
 
It can be shown that the form $\omega_{\pm}$ (\ref{KY_form}) is the generalized conformal Killing-Yano form with respect to the torsion:
\begin{equation}\label{torsion}
T_{\pm}=\frac{a}{\Delta^{3/4}_0}\left(r_1+r_2\pm2\chi(r)\right)\left(\sin{\theta}e^{0}\wedge e^{1}\wedge  e^3\mp\frac{\sqrt{X}\cos{\theta}}{\chi(r)}e^{0}\wedge e^{2}\wedge  e^3\right),
\end{equation}
where for simplicity we denote $r_i=r+2ms^2_i$, $i=1,2$ and $\chi(r)=\sqrt{r_1r_2}$. For the particular case when all the charges are equal ($\delta_1=\delta_2$), namely for the dyonic black hole with equal electric and magnetic charges as well as for the Kerr case the torsion $T_{-}$ turns to be zero, whereas $T_{+}$ does not. Therefore, only the Killing-Yano tensors $\omega_{-}$ and $f_{-}$ correspond to the standard limit, and the tensors $\omega_{+}$ and $f_{+}$ represent a ``new'' physical system. It was pointed out in \cite{Houri_JHEP10} that similar peculiarity takes place for the Kerr-Sen solution.

If the torsion is taken into consideration the Dirac operator $\hat{D}=\hat{\gamma}^{\mu}(\partial_{\mu}+\Gamma_{\mu})$ should be repalced by following operator:
\begin{equation}\label{D_mod}
\hat{\cal D}=\hat{D}-\frac{1}{24}T_{ABC}\hat{\gamma}^{A}\hat{\gamma}^{B}\hat{\gamma}^{C},
\end{equation}
where $T_{ABC}$ are the Lorentzian components of the torsion form (\ref{torsion}) and $\hat{\gamma}^{A}$ are the Lorentzian gamma matrices (\ref{gamma_matr}). The second term in (\ref{D_mod}) can be written in a space-time basis as well. The operator $\hat{\cal D}$ may be called the generalized Dirac operator.

Having defined the generalized Dirac operator (\ref{D_mod}) we write the generalized (modified) Dirac equation as follows:
\begin{equation}\label{D_eq_mod}
\hat{\cal D}\Psi+\mu_e\Psi=0.
\end{equation}
Since the torsion contribution into the modified Dirac operator (\ref{D_mod}) is just an algebraic term it means that the modified Dirac equation (\ref{D_eq_mod}) can be derived via standard variational procedure from the massive counterpart of the modified Dirac action (\ref{GD_action}) given in the Appendix \ref{app_A}.

Now we are able to write the explicit form for the modified Dirac equation (\ref{D_mod}) in the Einstein frame (\ref{tetrad_0}):
\begin{multline}
\left[\hat{\gamma}^{0}\frac{1}{\De^{1/4}_{0}\sqrt{X}}\left(F(r)\partial_{t}+a\partial_{\vp}\right)+\hat{\gamma}^{1}\frac{\sqrt{X}}{\De^{1/4}_0}\left(\partial_r+\frac{1}{2\De^{1/4}_0\sqrt{X}}\left(\De^{1/4}_{0}\sqrt{X}\right)'\right)\right.\\\left.+\hat{\gamma}^{2}\frac{1}{\De^{1/4}_0}\left(\partial_{\th}+\frac{1}{2\De^{1/4}_0\sin{\th}}\left(\De^{1/4}_{0}\sin{\th}\right)_{,\th}\right)-\hat{\gamma}^{3}\frac{1}{\De^{1/4}_0}\left(a\sin{\th}\partial_t+\frac{1}{\sin\th}\partial_{\vp}\right)\mp\right.\\\left.\hat{\gamma}^{3}\hat{\gamma}^{0}\hat{\gamma}^{1}\frac{a\chi(r)}{2\De^{3/4}_0}\sin{\th}\pm\hat{\gamma}^{3}\hat{\gamma}^{0}\hat{\gamma}^{2}\frac{a\sqrt{X}(r_1+r_2)}{4\chi(r)\De^{3/4}_{0}}\cos{\th}+\mu_e\hat{I}_{4}\right]\Psi=0.\label{D_mod_exp}
\end{multline}
To obtain the separation of variables the Dirac spinor $\Psi$ is taken in the following form:
\begin{equation}\label{D_sp}
\Psi=\begin{pmatrix}
F_{1}\\  F_{2}\\G_{1} \\ G_{2}\\
\end{pmatrix}.
\end{equation} 
Therefore equations for the components of the spinor $\Psi$ can be written as follows:
\begin{eqnarray}\label{D0}
\frac{i\sqrt{X}}{\De^{1/4}_0}\left(\hat{\cal D}_{-}+g_{\pm}(r,\th)\right)G_1+\frac{i}{\De^{1/4}_0}\left(\hat{\cal L}_{-}+f_{\pm}(r,\th)\right)G_2+\mu_{e}F_{1}=0;\\
\frac{i}{\De^{1/4}_0}\left(\hat{\cal L}_{+}+f_{\pm}(r,\th)\right)G_1-\frac{i\sqrt{X}}{\De^{1/4}_0}\left(\hat{\cal D}_{+}+g_{\pm}(r,\th)\right)G_2+\mu_{e}F_{2}=0;\label{D1}\\
-\frac{i\sqrt{X}}{\De^{1/4}_0}\left(\hat{\cal D}_{+}+g^{*}_{\pm}(r,\th)\right)F_1-\frac{i}{\De^{1/4}_0}\left(\hat{\cal L}_{-}+f^{*}_{\pm}(r,\th)\right)F_2+\mu_{e}G_{1}=0;\label{D2}\\
-\frac{i}{\De^{1/4}_0}\left(\hat{\cal L}_{+}+f^{*}_{\pm}(r,\th)\right)F_1+\frac{i\sqrt{X}}{\De^{1/4}_0}\left(\hat{\cal D}_{-}+g^{*}_{\pm}(r,\th)\right)F_2+\mu_{e}G_{2}=0,\label{D3}
\end{eqnarray}
where the following operators are introduced:
\begin{eqnarray}
\hat{{\cal D}}_{\pm}=\partial_{r}\mp\frac{1}{X}\left(F(r)\partial_{t}+a\partial_{\vp}\right),\label{D_op}\\\hat{{\cal L}}_{\pm}=\partial_{\th}\mp i\left[a\sin{\th}\partial_{t}+\frac{1}{\sin{\th}}\partial_{\vp}\right],\label{L_op}
\end{eqnarray}
and the functions $g_{\pm}(r,\th)$ and $f_{\pm}(r,\th)$ are defined as follows:
\begin{equation}
g_{\pm}(r,\th)=\frac{r-m}{2X}+\frac{r_1+r_2}{4\chi\rho^{*}_{\pm}},\quad f_{\pm}(r,\th)=\frac{\cot{\th}}{2}\pm\frac{ia\sin{\th}}{2\rho^{*}_{\pm}}.
\end{equation}
Here $\rho_{\pm}=\chi(r)\pm ia\cos{\th}$ and asterisk means complex conjugation, it is obvious that $\Delta^{1/2}_{0}=\rho_{\pm}\rho^{*}_{\pm}$. We point out here that $\rho_{+}$ and $\rho_{-}$ are complex conjugates to each other, and consequently the same is true for the functions $g_{\pm}(r,\th)$ and $f_{\pm}(r,\th)$, but we emphasize that indices $\pm$ for the functions $\rho$ and consequently for $g$ and $f$ are introduced to distinguish the cases which correspond to two values of the torsion $T_{\pm}$ (\ref{torsion}). It should be noted that separability of the radial $r$ and angular over $\theta$ parts in the system (\ref{D0})-(\ref{D3}) is closely related to a specific relation for derivatives of the functions $g$ and $f$, namely $\left(g_{\pm}\right)_{,\theta}=f'_{\pm}$. 

Further simplification of the equations (\ref{D0})-(\ref{D3}) can be achieved if the Dirac spinor $\Psi$ is taken in the form:
\begin{equation}\label{D_sp2}
\Psi=\frac{1}{\sqrt[4]{X}\sqrt{\sin{\th}}}\begin{pmatrix}\rho^{-1/2}_{\pm}\bar{F}_1\\\rho^{-1/2}_{\pm}\bar{F}_2 \\\left(\rho^{*}_{\pm}\right)^{-1/2}\bar{G}_1\\\left(\rho^{*}_{\pm}\right)^{-1/2}\bar{G}_2
\end{pmatrix},
\end{equation}
 Substituting the components of the Dirac spinor (\ref{D_sp2}) into the system (\ref{D0})-(\ref{D3}) and after some transformations we can rewrite the upper system in the form:  
\begin{eqnarray}
\sqrt{X}\hat{{\cal D}}_{-}\bar{G}_{1}+\hat{{\cal L}}_{-}\bar{G}_{2}-i\mu_{e}\rho^{*}_{\pm}\bar{F}_{1}=0;\label{D_13}\\\hat{{\cal L}}_{+}\bar{G}_{1}-\sqrt{X}\hat{{\cal D}}_{+}\bar{G}_{2}-i\mu_{e}\rho^{*}_{\pm}\bar{F}_{2}=0;\label{D_23}\\\sqrt{X}\hat{{\cal D}}_{+}\bar{F}_{1}+\hat{{\cal L}}_{-}\bar{F}_{2}+i\mu_{e}\rho_{\pm}\bar{G}_{1}=0;\label{D_33}\\\hat{{\cal L}}_{+}\bar{F}_{1}-\sqrt{X}\hat{{\cal D}}_{-}\bar{F}_{2}+i\mu_{e}\rho_{\pm}\bar{G}_{2}=0.\label{D_43}
\end{eqnarray}
The above system of equations can be represented in an equivalent matrix form:
\begin{equation}
\begin{pmatrix}
-i\mu_{e}\rho^{*}_{\pm} & 0 & \sqrt{X}\hat{\cal D}_{-} & \hat{\cal L}_{-}\\0 & -i\mu_{e}\rho^{*}_{\pm} & \hat{\cal L}_{+} & -\sqrt{X}\hat{\cal D}_{+}\\\sqrt{X}\hat{\cal D}_{+} & \hat{\cal L}_{-} & i\mu_{e}\rho_{\pm} & 0 \\ \hat{\cal L}_{+} & -\sqrt{X}\hat{\cal D}_{-} & 0 & i\mu_{e}\rho_{\pm}
\end{pmatrix}
\begin{pmatrix}
\bar{F}_{1} \\ \bar{F}_{2} \\ \bar{G}_{1} \\ \bar{G}_{2}
\end{pmatrix}=0.\label{D_matr}
\end{equation}
Finally, to separate variables in the system of equations (\ref{D_matr})  we extract an exponential factor which depends on $t$ and $\vp$ and use a specific ansatz for the part which depends on $r$ and $\th$, namely we write:
\begin{equation}\label{chan_wf}
\begin{pmatrix}
\bar{F}_{1} \\ \bar{F}_{2} \\ \bar{G}_{1} \\ \bar{G}_{2}
\end{pmatrix}=e^{i(\omega t+k\vp)}\begin{pmatrix}
{\cal R}_{2}(r){\cal S}_{1}(\th) \\ {\cal R}_{1}(r){\cal S}_{2}(\th) \\ {\cal R}_{1}(r){\cal S}_{1}(\th) \\ {\cal R}_{2}(r){\cal S}_{2}(\th)
\end{pmatrix}.
\end{equation}
Having used the upper ansatz we decouple the system (\ref{D_matr}) on two systems and achieve separation of variables. 
Therefore, after some algebraic transformations we arrive at the following decoupled systems of equations for radial and angular parts:
\begin{equation}
\begin{pmatrix}\label{rad_sys}
\sqrt{X}\hat{\tilde{\cal D}}_{-} & -i\mu_{e}\chi(r)-\la \\i\mu_{e}\chi(r)-\la & \sqrt{X}\hat{\tilde{\cal D}}_{+}
\end{pmatrix}\begin{pmatrix} {\cal R}_{1}(r)\\ {\cal R}_{2}(r)\end{pmatrix}=0,
\end{equation}
\begin{equation}
\begin{pmatrix}\label{ang_sys}
\hat{\tilde{\cal L}}_{+} & -\la\mp\mu_{e}a\cos{\th} \\ \la\mp\mu_{e}a\cos{\th} & \hat{\tilde{\cal L}}_{-}\end{pmatrix}\begin{pmatrix} {\cal S}_1(\th) \\{\cal S}_2(\th)\end{pmatrix}=0,
\end{equation}
where $\la$ is  the separation constant and
\begin{eqnarray}
{\hat{\tilde{\cal D}}}_{\pm}=\partial_{r}\mp\frac{i}{X(r)}\left(\om F(r)+ka\right),\\ {\hat{\tilde{\cal L}}}_{\pm}=\partial_{\th}\pm\left(a\om\sin{\th}+\frac{k}{\sin{\th}}\right).
\end{eqnarray}
We analyze the systems (\ref{rad_sys}) and (\ref{ang_sys}) in more details in the fifth and sixth sections, paying considerable attention to the massless case.

\section{The Dirac equation in the String frame}\label{section_4}
Apart of the Einstein frame considered in the previous sections there is another frame of almost equal importance, namely the String frame. We do not discuss properties or peculiarites of both of them, but we just point out that they are conformally related, namely we write:
\begin{equation}\label{conf_rel}
d\tilde{s}^2=e^{\phi}ds^2 \quad \Leftrightarrow \quad \tilde{g}_{\mu\nu}=e^{\phi}g_{\mu\nu},
\end{equation}
where notations with and without tilde correspond to the String and the Einstein frames respectively. The conformal factor $e^{\phi}$ takes the following form \cite{Chong_NPB05}:
\begin{equation}\label{dilat}
e^{\phi}=\frac{(r+2ms^2_1)^2+a^2\cos^2{\theta}}{(r+2ms^2_1)(r+2ms^2_2)+a^2\cos^2{\theta}}.
\end{equation}
Taking into account the relations (\ref{conf_rel}), (\ref{dilat}) and (\ref{tetrad_0}) we obtain the Boyer-Lindquist tetrad in the String frame:
\begin{eqnarray}\label{tetrad_SF}
\nonumber \tilde{e}^{0}=\frac{R_1\sqrt{X}}{\De^{1/2}_0}\left(dt-a\sin^2\th d\vp\right),\quad \tilde{e}^{1}=\frac{R_1}{\sqrt{X}}dr;\quad \tilde{e}^2 =R_{1}d\th, \\ \tilde{e}^{3}=\frac{R_1\sin{\th}}{\De^{1/2}_0}\left(adt-\left((r+2ms^2_{1})(r+2ms^2_{2})+a^2\right)d\vp\right),
\end{eqnarray}
where $R^2_1=(r+2ms^2_1)^2+a^2\cos^2{\theta}$. Spin-connection $1$-forms ${{\tilde{\omega}}^{A}}_{B}$ for the String frame can be derived directly as above, or  the rescaling relation between the frames can be utilized. Explicit expressions for the spin-connection $1$-forms and relations for spin-connection components in conformally related frames are given in the Appendix \ref{app_A}. Now we write the explicit relation for the contracted product $\hat{\tilde{\gamma}}^{\mu}\tilde{\Gamma}_{\mu}$:
\begin{multline}
\hat{\tilde{\gamma}}^{\mu}\tilde{\Gamma}_{\mu}=\frac{1}{2}\left(\frac{\De^{1/2}_{0}}{R^4_1}\left(\frac{R^3_1\sqrt{X}}{\De^{1/2}_0}\right)'\hat{\gamma}^{1}+\frac{\De^{1/2}_{0}}{R^4_1\sin{\th}}\left(\frac{R^3_1\sin{\th}}{\De^{1/2}_{0}}\right)_{,\th}\hat{\gamma}^{2}\right.\\\left.+\frac{aF'}{2R_1\De^{1/2}_{0}}\sin{\th}\hat{\gamma}^{3}\hat{\gamma}^{0}\hat{\gamma}^{1}-\frac{a\sqrt{X}}{R_1\De^{1/2}_0}\cos{\th}\hat{\gamma}^{3}\hat{\gamma}^{0}\hat{\gamma}^{2}\right).
\end{multline}

The generlaized conformal Killing-Yano tensor for the Kerr-Sen background in the String frame was derived in a different way \cite{Houri_JHEP10}, if the torsion was associated directly with the Kalb-Ramond field or in the dual description with the Hodge-dual to the axion field. On the other hand, it was proven that the generalized conformal Killing-Yano and the torsion forms (or tensors) undergo a simple rescaling if a confrormal rescaling of the background metric is performed \cite{Houri_JHEP10,Houri_CQG10}. Two ways to derive torsion and generalized conformal Killing-Yano forms for the Kerr-Sen background show that the resulting values are not conformally related to their counterparts obtained in a different manner, therefore we can conjecture that these two ways give rise to different physical systems on the same background. Morover in the Appendix \ref{app_A} below we show that the massless torsion-modified Dirac action is conformally invariant if a proper rescaling of spinors is performed, this fact generlaizes the well-known statement about the conformal invariance of the stanadard massless Dirac action. The straightforward consequence of this statement is the conclusion about the conformal invariance of the generalized Dirac equation. This fact can be considered as an additional confirmation of our conjection that we actually deal with two different Dirac fields, even though they look very similar.

Now we can consider the generalized Killing-Yano tensors in the String frame and show that they satisfy the equation completely identical to (\ref{KY_eq_form}), but with correspondingly redefined covariant derivative $\tilde{\nabla}\tilde{\omega}$ and co-derivative $\delta{\tilde{\omega}}$ consistent with the rescaled metric. The generalized conformal Killing-Yano tensor $\tilde{\omega}$ (GCKYT) and the generalized Killing-Yano tensor $\tilde{f}$ (GKYT) in the String frame take the form as follows:
\begin{eqnarray}\label{GKYT_SF}
\tilde{\omega}_{\pm}=(r+2ms^2_1)\tilde{e}^{0}\wedge\tilde{e}^{1}\pm a\cos{\theta}\tilde{e}^{2}\wedge\tilde{e}^{3};\\\tilde{f}_{\pm}=-(r+2ms^2_1)\tilde{e}^{2}\wedge\tilde{e}^{3}\pm a\cos{\theta}\tilde{e}^{0}\wedge\tilde{e}^{1}.
\end{eqnarray}
Similarly as in the Einstein frame the forms $\tilde{\omega}_{\pm}$ and $\tilde{f}_{\pm}$ are supposed to be $T$-closed and $T$-co-closed respectively with a new torsion form $\tilde{T}_{\pm}$. The latter restraints allow us to obtain explicit expression for the torsion form in the String frame, namely we write:
\begin{multline}\label{torsion_SF}
\tilde{T}_{\pm}=\frac{a\sin{\th}}{R_1}\left(\frac{F'}{\De^{1/2}_{0}}\pm\frac{2(r+2ms^2_1)}{R^2_1}\right)\tilde{e}^{0}\wedge\tilde{e}^{1}\wedge\tilde{e}^{3}-\frac{2a\sqrt{X}\cos{\th}}{R_1}\left(\frac{1}{\De^{1/2}_{0}}\pm\frac{1}{R^2_1}\right)\tilde{e}^{0}\wedge\tilde{e}^{2}\wedge\tilde{e}^{3}.
\end{multline} 
The torsion $\tilde{T}_{-}$ coincides up to a sign with the Kalb-Ramond field calculated in the String frame and the explicit form of the latter one is given in the Appendix B. The difference up to the total sign can be explained by the fact  that both of the Killing-Yano tensors $\tilde{\omega}_{\pm}$ and $\tilde{f}_{\pm}$ are defined up to a total sign, and if they change the sign the torsion does it as well. The $\pm$ subscripts in the torsion $\tilde{T}$ have the same meaning as for the torsion (\ref{torsion}) in the Einstein frame, namely the torsion $\tilde{T}_{-}$ equals to zero for the equal charges dyonic black hole and for the Kerr solution, whereas the torsion $\tilde{T}_{+}$ remains nontrivial, but again both of them give rise to the separable Dirac equations.  For the Kerr limit both frames are actually reduced to the standard Boyer-Lindquist frame and consequently for the Killing-Yano tensors we have $\omega_{\pm}=\tilde{\omega}_{\pm}$ and $f_{\pm}=\tilde{f}_{\pm}$ as it is expected. Similar conclusion can be made for the equal charges dyonic black hole. 

The modified Dirac equation in the String frame is constructed in the same way in the Einstein one, namely we use  the modified Dirac operator defined by (\ref{D_mod}), but the standard Dirac operator should be written now in the frame (\ref{tetrad_SF}) and the torsion field components are taken from (\ref{torsion_SF}).
Thus we write:
\begin{multline}\label{Dirac_SF}
\left[\hat{\gamma}^{0}\frac{1}{R_1\sqrt{X}}\left(F\partial_{t}+a\partial_{\vp}\right)+\hat{\gamma}^{1}\left(\frac{\sqrt{X}}{R_1}\partial_{r}+\frac{\De^{1/2}_0}{2R^4_1}\left(\frac{R^3_1\sqrt{X}}{\De^{1/2}_0}\right)'\right)+\right.\\\left.\hat{\gamma}^2\left(\frac{1}{R_1}\partial_{\theta}+\frac{\De^{1/2}_0}{2R^4_1\sin{\th}}\left(\frac{R^3_1\sin{\th}}{\De^{1/2}_{0}}\right)_{,\th}\right)-\hat{\gamma}^{3}\frac{1}{R_1}\left(a\sin{\th}\partial_{t}+\frac{1}{\sin{\th}}\partial_{\vp}\right)\right.\\\left.\mp\frac{ar_1\sin{\th}}{2R^3_1}\hat{\gamma}^{3}\hat{\gamma}^{0}\hat{\gamma}^{1}\pm\frac{a\sqrt{X}\cos{\th}}{2R^3_1}\hat{\gamma}^{3}\hat{\gamma}^{0}\hat{\gamma}^{2}+\mu_{e}\hat{I}_{4}\right]\tilde{\Psi}_{1}=0.
\end{multline}
Since for the Dirac equation in the String frame (\ref{Dirac_SF}) we use the same procedure as in the previous section, here we focus on the key differences. Namely, the Dirac spinor $\tilde{\Psi}_{1}$ in the equation (\ref{Dirac_SF}) is taken in the form:
\begin{equation}\label{Dir_sp_SF}
\tilde{\Psi}_{1}=e^{i(\omega t+k\vp)}\frac{\De^{1/4}_0}{R_1\sqrt[4]{X}\sqrt{\sin{\th}}}\begin{pmatrix}\rho_{(1)\pm}^{-1/2}\tilde{F}_1\\\rho^{-1/2}_{(1)\pm}\tilde{F}_2 \\\left(\rho^{*}_{(1)\pm}\right)^{-1/2}\tilde{G}_1\\\left(\rho^{*}_{(1)\pm}\right)^{-1/2}\tilde{G}_2
\end{pmatrix},
\end{equation}
where $\rho_{(1)\pm}=r_1\pm i\cos{\th}$ and $\rho^{*}_{(1)\pm}$ is its complex conjugate. We point out that $\rho_{(1)+}$ and $\rho_{(1)-}$ are mutually complex conjugate functions, but the indices $\pm$ for these functions show that they correspond to two different torsion forms $\tilde{T}_{\pm}$, similarly to the functions $\rho_{\pm}$ in the previous section. Since the time-$t$ and angular-$\varphi$ dependences are already explicitly given in the spinor (\ref{Dir_sp_SF}), therefore its components $\tilde{F}_i$ and $\tilde{G}_i$ ($i=1,2$) depend on the radial $r$ and angular $\theta$ variables only.

Substituting the spinor (\ref{Dir_sp_SF}) into the equation (\ref{Dirac_SF}) and after some transformations we arrive at the coupled system of equations analogous to the system (\ref{D_matr}), but with corresponding replacement of the functions $\rho_{\pm}$ and $\rho^{*}_{\pm}$ by the functions $\rho_{(1)\pm}$ and $\rho^{*}_{(1)\pm}$ respectively in the diagonal elements of the matrix and the operators $\hat{\cal{D}}_{\pm}$ and $\hat{\cal{L}}_{\pm}$ should be replaced by the operators $\hat{\tilde{\cal{D}}}_{\pm}$ and $\hat{\tilde{\cal{L}}}_{\pm}$ since we have already decoupled dependences over $t$ and $\vp$. Further procedure to completely decouple the dependence over $r$  and $\th$ is again analogous to what is done in the previous section, namely for the spinor with the components $\tilde{F}_i$ and $\tilde{G}_i$ ($i=1,2$) we use the same ansatz as in (\ref{chan_wf}) and after some transformations we obtain two decoupled systems of equations for the radial and angular parts which are similar to the systems (\ref{rad_sys}) and (\ref{ang_sys}) correspondingly. To be more precise, the angular system is completely the same as (\ref{ang_sys}) and in the radial system (\ref{rad_sys}) the function $\chi$ should be replaced by $r_1$.  For massless fermions the corresponding systems will be the same in both frames.

Conformal invariance of the generalized Dirac action (\ref{GD_action}) shown in the Appendix \ref{app_A}, brings the conclusion about conformal invariance of the Dirac equation if the Dirac spinor is rescaled properly. But in the massive case the mass term is also rescaled and it turns to be a function of coordinates what might spoil separability of the system. 

In order not to lose separability under the conformal rescaling now we consider the massless case only and show that under the conformal rescaling of the massless analog of the equation (\ref{Dirac_SF}) we will not arrive at massless form the equation (\ref{D_mod_exp}), but obtain a conformal cousin of the equation (\ref{Dirac_SF}). Here below we describe just the keypoints, because there are many similarities with the transformations made above. Similar analysis will be made for the equation (\ref{D_mod_exp}) as well.

First of all we transform the contracted torsion term $\tilde{T}_{ABC}\hat{\gamma}^A\hat{\gamma}^B\hat{\gamma}^C/24$ for the torsion $\tilde{T}_{\pm}$ (\ref{torsion_SF}) when the String frame values are transformed into the Einstein frame ones. We write:
\begin{multline}
\frac{1}{24}T_{\mu\nu\la}\hat{\gamma}^{\mu}\hat{\gamma}^{\nu}\hat{\gamma}^{\la}=\frac{1}{24}e^{\phi/2}\tilde{T}_{\mu\nu\la}\hat{\tilde{\gamma}}^{\mu}\hat{\tilde{\gamma}}^{\nu}\hat{\tilde{\gamma}}^{\la}=\frac{a\sin{\th}}{4\De^{1/4}_{0}}\left(\frac{F'}{\De^{1/2}_{0}}\pm\right.\\\left.\frac{2(r+2ms^2_1)}{R^2_1}\right)\hat{\gamma}^{3}\hat{\gamma}^{0}\hat{\gamma}^{1}-\frac{a\sqrt{X}\cos{\th}}{2\De^{1/4}_{0}}\left(\frac{1}{\De^{1/2}_{0}}\pm\frac{1}{R^2_1}\right)\hat{\gamma}^{3}\hat{\gamma}^{0}\hat{\gamma}^{2}.
\end{multline}
Taking into account the written above contracted torsion term we write conformally rescaled massless Dirac equation (\ref{Dirac_SF}) as follows:
\begin{multline}
\left[\hat{\gamma}^{0}\frac{1}{\De^{1/4}_{0}\sqrt{X}}\left(F(r)\partial_{t}+a\partial_{\vp}\right)+\hat{\gamma}^{1}\frac{\sqrt{X}}{\De^{1/4}_0}\left(\partial_r+\frac{1}{2\De^{1/4}_0\sqrt{X}}\left(\De^{1/4}_{0}\sqrt{X}\right)'\right)+\right.\\\left.\hat{\gamma}^{2}\frac{1}{\De^{1/4}_0}\left(\partial_{\th}+\frac{1}{2\De^{1/4}_0\sin{\th}}\left(\De^{1/4}_{0}\sin{\th}\right)_{,\th}\right)-\hat{\gamma}^{3}\frac{1}{\De^{1/4}_0}\left(a\sin{\th}\partial_t+\frac{1}{\sin\th}\partial_{\vp}\right)\right.\\\left.\mp\frac{ar_{1}\sin{\th}}{2\De^{1/4}_{0}R^2_1}\hat{\gamma}^{3}\hat{\gamma}^{0}\hat{\gamma}^{1}\pm\frac{a\sqrt{X}\cos{\th}}{2\De^{1/4}_{0}R^2_1}\hat{\gamma}^{3}\hat{\gamma}^{0}\hat{\gamma}^{2}\right]\Psi_{1}=0. \label{Dir_S_to_E}
\end{multline}
To perform separation variables in the equation (\ref{Dir_S_to_E}) we take the following ansatz for the spinor wave-function $\Psi_{1}$:
\begin{equation}\label{Psi_S_to_E}
\Psi_{1}=e^{i(\omega t+k\vp)}\frac{R^{1/2}_{1}}{\De^{1/8}_0\sqrt[4]{X}\sqrt{\sin{\th}}}\begin{pmatrix}\rho_{(1)\pm}^{-1/2}F_1\\\rho^{-1/2}_{(1)\pm}F_2 \\\left(\rho^{*}_{(1)\pm}\right)^{-1/2}G_1\\\left(\rho^{*}_{(1)\pm}\right)^{-1/2}G_2
\end{pmatrix}.
\end{equation}
Now it is easy to check that the rescaling relation for the spinors (\ref{Psi_S_to_E}) and (\ref{Dir_sp_SF}), namely we obtain:
\begin{equation}
\Psi_1=\frac{R^{3/2}_1}{\De^{3/8}_{0}}\tilde{\Psi}_{1},
\end{equation}
as it should be according to the rescaling relation (\ref{spinor_resc}). 

Similarly, rescaling the torsion $T_{\pm}$ (\ref{torsion}) and after necessary transformations we rewrite the Dirac equation (\ref{D_mod_exp}) (massless case) now in the String frame:
\begin{multline}\label{Dirac_E_to_S}
\left[\hat{\gamma}^{0}\frac{1}{R_1\sqrt{X}}\left(F\partial_{t}+a\partial_{\vp}\right)+\hat{\gamma}^{1}\left(\frac{\sqrt{X}}{R_1}\partial_{r}+\frac{\De^{1/2}_0}{2R^4_1}\left(\frac{R^3_1\sqrt{X}}{\De^{1/2}_0}\right)'\right)+\right.\\\left.\hat{\gamma}^2\left(\frac{1}{R_1}\partial_{\theta}+\frac{\De^{1/2}_0}{2R^4_1\sin{\th}}\left(\frac{R^3_1\sin{\th}}{\De^{1/2}_{0}}\right)_{,\th}\right)-\hat{\gamma}^{3}\frac{1}{R_1}\left(a\sin{\th}\partial_{t}+\frac{1}{\sin{\th}}\partial_{\vp}\right)\right.\\\left.\mp\frac{a\chi\sin{\th}}{2R_1\De^{1/2}_{0}}\hat{\gamma}^{3}\hat{\gamma}^{0}\hat{\gamma}^{1}\pm\frac{aF'\sqrt{X}\cos{\th}}{2R_1\De^{1/2}_{0}\chi}\hat{\gamma}^{3}\hat{\gamma}^{0}\hat{\gamma}^{2}\right]\tilde{\Psi}=0.
\end{multline}
To perform separation of variables we use ansatz:
\begin{equation}\label{Psi_E_to_S}
\tilde{\Psi}=e^{i(\omega t+k\vp)}\frac{\De^{3/8}_{0}}{R^{3/2}_1\sqrt[4]{X}\sqrt{\sin{\th}}}\begin{pmatrix}\rho^{-1/2}_{\pm}\bar{F}_1\\\rho^{-1/2}_{\pm}\bar{F}_2 \\\left(\rho^{*}_{\pm}\right)^{-1/2}\bar{G}_1\\\left(\rho^{*}_{\pm}\right)^{-1/2}\bar{G}_2
\end{pmatrix}.
\end{equation}
Now we can easily check that the spinor wave-functions (\ref{Psi_E_to_S}) and (\ref{D_sp2}) are related by:
\begin{equation}\label{psi_ES_rel}
\tilde{\Psi}=\frac{\De^{3/8}_{0}}{R^{3/2}_1}\Psi,
\end{equation}
as it should be.

\section{Radial and angular wavefunctions for the massless case}

We have noted above that in the massless case $\mu_e=0$ the decoupled systems of equations (\ref{rad_sys}) and (\ref{ang_sys}) for radial and angular components are completely the same for both types of torsions and frames we consider. The massless case is simpler for analysis at the same it reveals the main features, caused by the space-time geometry. We also point out that for the massless case the system of equations (\ref{D0})-(\ref{D3}) decouples on two systems of equations (\ref{D0})-(\ref{D1}) and (\ref{D2})-(\ref{D3}) for the lower and upper components of the Dirac spinor respectively and it corresponds to the parts of different chiralities. 

Considering the system (\ref{rad_sys}) we decouple the equations for the functions ${\cal R}_{1}(r)$ and ${\cal R}_{2}(r)$:
 \begin{eqnarray}\label{R1_m0}
{\cal R}''_{1}+\frac{X'}{2X}{\cal R}'_{1}+\frac{1}{X}\left.\Big(i\omega F'+\frac{(\omega F+ka)}{X}\left(\omega F+ka-\frac{i}{2}X'\right)-\lambda^2\right){\cal R}_{1}=0;\\\label{R2_m0}
{\cal R}''_{2}+\frac{X'}{2X}{\cal R}'_2+\frac{1}{X}\left.\Big(-i\omega F'+\frac{(\omega F+ka)}{X}\left(\omega F+ka+\frac{i}{2}X'\right)-\lambda^2\right){\cal R}_{2}=0.
\end{eqnarray}
It should be noted that the general structure of the radial equations for the massless case is very similar to their massless counterparts in the standard Kerr case. It can be explained by the fact that the singularities of the equations (\ref{R1_m0}) and (\ref{R2_m0}) are completely defined by the function $X(r)$ which is the same as in the Kerr case.  

To analyze the radial Dirac equations (\ref{R1_m0}) and (\ref{R2_m0}) in more details we rewrite them in a more tractable form. First of all we note that the function $X(r)$ can be rewritten in the form $X(r)=(r-r_+)(r-r_{-})$, where $r_{\pm}=m\pm\sqrt{m^2-a^2}$ are the horizons of the black hole and in particular $r_+$ is its outer or event horizon. 
After some algebraic transformations the radial equations (\ref{R1_m0}) and (\ref{R2_m0}) can be written in the form:
\begin{multline}\label{rad_Ri}
{\cal R}''_{j}+\frac{1}{2}\left(\frac{1}{r-r_+}+\frac{1}{r-r_{-}}\right){\cal R}'_{j}+\left(\frac{A_j}{r-r_{+}}+\frac{B_j}{r-r_{-}}+\frac{C_j}{(r-r_{+})^2}+\frac{D_j}{(r-r_{-})^2}+\omega^2\right){\cal R}_j=0,
\end{multline}
where $j=1,2$ and the coefficients $A_j$, $B_j$, $C_j$ and $D_j$ are as follows:
\begin{gather}
\label{A_j} A_{j}=\pm i\om\left(B_{+}-\frac{1}{2}\right)+2\om A_{+}+\frac{2A_{+}A_{-}\mp i(A_{+}+A_{-})/2-\la^2}{r_{+}-r_{-}};\\\label{B_j}
B_{j}=\pm i\om\left(B_{-}-\frac{1}{2}\right)+2\om A_{-}-\frac{2A_{+}A_{-}\mp i(A_{+}+A_{-})/2-\la^2}{r_{+}-r_{-}};\\\label{CD_j}
C_{j}=A_{+}\left(A_{+}\mp\frac{i}{2}\right);\quad
D_{j}=A_{-}\left(A_{-}\mp\frac{i}{2}\right);
\end{gather}
and it should be pointed out that upper and lower sings in the written above relations correspond to $j=1$ and $j=2$ respectively. For simplicity of the written above relations we have used the following notations:
\begin{eqnarray}
\nonumber A_{\pm}=\pm\frac{2m(1+s^2_1+s^2_2)\om r_{\pm}+4\om m^2s^2_1s^2_2+ka}{r_{+}-r_{-}}, \quad B_{\pm}=\pm\frac{2(r_{\pm}+m(s^2_1+s^2_2))}{r_{+}-r_{-}}.
\end{eqnarray}
The equation (\ref{rad_Ri}) has singular points at the horizons $r=r_{\pm}$ and at the infinity $r\to\infty$. To reduce the equation (\ref{rad_Ri}) to the canonical form we  perform the transformation of coordinates:
\begin{equation}\label{x_r}
x=\frac{r-r_{-}}{r_{+}-r_{-}},
\end{equation}
 and as a consequence the singular points $r=r_{-}$ and $r=r_{+}$ are transformed into $x=0$ and $x=1$ respectively. The equation (\ref{rad_Ri}) can be simplified, namely the terms proportional to $\sim 1/(r-r_{+})^2$, $\sim 1/(r-r_{-})^2$ and $\sim\om^2$ can be removed from the equation (\ref{rad_Ri}) if one specifies a structure of the radial wave-function. Taking into account the transformation (\ref{x_r}) we write the radial wave-function ${\cal R}_i$ as follows:
 \begin{equation}\label{ansatz_Ri}
 {\cal R}_j(x)=e^{\kappa x}x^{\alpha_j}(x-1)^{\beta_j}\tilde{{\cal R}}_{j}(x),
 \end{equation}
where the coefficients $\kappa$, $\alpha_j$ and $\beta_j$ satisfy the following relations:
\begin{equation}\label{cond_ab}
\kappa=\pm i(r_{+}-r_{-})\om, \quad \alpha_j\left(\alpha_j-\frac{1}{2}\right)+D_{j}=0, \quad \beta_j\left(\beta_j-\frac{1}{2}\right)+C_{j}=0.
\end{equation}
We note that upper and lower signs for the parameter $\kappa$ correspond to ingoing and outgoing particles respectively. For a particular problem one or two signs can be chosen. The equations (\ref{cond_ab}) allow to obtain explicit relations for the exponents $\al_j$ and $\be_j$. Since the equations (\ref{cond_ab}) for $\al_j$ and $\be_j$ are quadratic it means that for any fixed value of $j$ we obtain two solutions. Namely, for the parameter $\al_j$ we derive:
\begin{equation}\label{al_coef}
\al_{j(1)}=\frac{1}{2}\mp iA_{-}, \quad \al_{j(2)}=\pm iA_{-}.
\end{equation}
For the parameter $\be_j$ we write:
\begin{equation}\label{be_coef}
\be_{j(1)}=\frac{1}{2}\mp iA_{+}, \quad \be_{j(2)}=\pm iA_{+}.
\end{equation}
Here we also point out that similarly to the relations (\ref{A_j})-(\ref{CD_j}) upper and lower signs in the relations (\ref{al_coef}) and (\ref{be_coef}) correspond to $j=1$ and $j=2$ respectively.
  
Finally, after all these transformation the equation (\ref{rad_Ri}) can be rewritten in the form:
\begin{equation}\label{rad_Ri_x}
\tilde{{\cal R}}''_{j}(x)+\left(2\kappa+\frac{2\alpha_{j}+1/2}{x}+\frac{2\beta_{j}+1/2}{x-1}\right)\tilde{{\cal R}}'_{j}(x)+\left(\frac{\mu_j}{x}+\frac{\nu_j}{x-1}\right)\tilde{{\cal R}}_{j}(x)=0,
\end{equation}
where the parameters $\mu_{j}$ and $\nu_{j}$ are defined as follows:
\begin{eqnarray}
\mu_{j}=2\alpha_j(\kappa-\beta_j)+\frac{1}{2}(\kappa-\alpha_{j}-\beta_{j})+(r_{+}-r_{-})B_{j}, \\\nu_{j}=2\beta_j(\kappa+\alpha_j)+\frac{1}{2}(\kappa+\alpha_{j}+\beta_{j})+(r_{+}-r_{-})A_{j}.
\end{eqnarray}
We note that in the equation (\ref{rad_Ri_x}) we again use prime $'$, to denote the derivative over coordinate $x$. This notation is used for simplicity, but it should not be confusing.

Now we consider the angular part of the wave function given by the system (\ref{ang_sys}), where we set $\mu_e=0$. It can be shown that  the equations for the angular part can be represented in the form:
\begin{equation}\label{ang_eq12}
\frac{\partial^2 {\cal S}_{j}}{\partial \theta^2}-\left(\left(a\om\sin{\theta}+\frac{k}{\sin{\th}}\right)^2\mp\left(a\om-\frac{k}{\sin^2{\th}}\right)\cos{\th}-\lambda^2\right){\cal S}_{j}=0,
\end{equation}
where similarly as above upper and lower signs correspond to $j=1$ and $j=2$ respectively. To rewrite the latter equation in more convenient form we make use of the following transformation of coordinates:
\begin{equation}\label{y_t}
y=\frac{1}{2}(1+\cos{\theta}).
\end{equation}
The transformation (\ref{y_t}) allows to rewrite the equation (\ref{ang_eq12}) as follows:
\begin{equation}\label{eq_sy}
\frac{\partial^2 {\cal S}_j}{\partial y^2}+\frac{1}{2}\left(\frac{1}{y}+\frac{1}{y-1}\right)\frac{\partial {\cal S}_j}{\partial y}+\left(\frac{\bar{A}_j}{y}+\frac{\bar{B}_j}{y-1}+\frac{\bar{C}_j}{y^2}+\frac{\bar{D}_j}{(y-1)^2}-4a^2\om^2\right){\cal S}_j=0,
\end{equation}
where we have:
\begin{gather*}
\bar{A}_j=-\frac{k^2}{2}-a\om(2k\mp 1)-\la^2, \quad \bar{B}_j=\frac{k^2}{2}+a\om(2k\pm 1)+\la^2, \\\bar{C}_j=-\frac{k(k\pm 1)}{4}, \quad \bar{D}_j=-\frac{k(k\mp 1)}{4},
\end{gather*}
and here similarly as earlier upper and lower signs correspond to $j=1$ and $j=2$ respectively. The general structure of the equation (\ref{eq_sy}) is completely identical to the equation (\ref{rad_Ri}) and it means that we can use an ansatz similar to (\ref{ansatz_Ri}) to simplify it, namely we write:
\begin{equation}\label{ansatz_Si}
{\cal S}_{j}(y)=e^{\bar{\kappa}y}y^{\bar{\alpha}_j}(y-1)^{\bar{\beta}_j}\bar{{\cal S}}_{j}(y),
\end{equation}
where the parameters $\bar{\kappa}$, $\bar{\alpha}_j$ and $\bar{\beta}_j$ fullfil the conditions:
\begin{equation}
\bar{\kappa}=\pm 2a\om, \quad \bar{\alpha}_j\left(\bar{\alpha}_j-\frac{1}{2}\right)+\bar{C}_{j}=0, \quad \bar{\beta}_j\left(\bar{\beta}_j-\frac{1}{2}\right)+\bar{D}_{j}=0.
\end{equation}
As a consequence, the equation (\ref{eq_sy}) can be represented in the form:
\begin{equation}\label{ang_eq_can}
\frac{\partial^2 \bar{{\cal S}}_j}{\partial y^2}+\left(2\bar{\kappa}+\frac{2\bar{\alpha}_{j}+1/2}{y}+\frac{2\bar{\beta}_{j}+1/2}{y-1}\right)\frac{\partial \bar{{\cal S}}_j}{\partial y}+\left(\frac{\bar{\mu}_j}{y}+\frac{\bar{\nu}_j}{y-1}\right)\bar{{\cal S}}_j=0,
\end{equation}
where the coefficients $\bar{\mu}_i$ and $\bar{\nu}_i$ are defined as follows:
\begin{eqnarray}
\bar{\mu}_{j}=2\bar{\alpha}_j(\bar{\kappa}-\bar{\beta}_j)+\frac{1}{2}(\bar{\kappa}-\bar{\alpha}_{j}-\bar{\beta}_{j})+\bar{A}_{j}, \\\bar{\nu}_{j}=2\bar{\beta}_j(\bar{\kappa}+\bar{\alpha}_j)+\frac{1}{2}(\bar{\kappa}+\bar{\alpha}_{j}+\bar{\beta}_{j})+\bar{B}_{j}.
\end{eqnarray}
The final equations for the radial (\ref{rad_Ri_x}) and angular (\ref{ang_eq_can}) components have similar structure, namely they are written in almost standard form for the confluent Heun equation. The confluent Heun equation is characterized by three singular points, namely, both our equations (\ref{rad_Ri_x}) and (\ref{ang_eq_can}) are rewritten in the form, where  the singularities are at the points $0$, $1$ and $\infty$. We also remark, that for the radial equation (\ref{rad_Ri_x}), the first two singularities correspond to the horizons and they are the so-called regular singularities. The last singularity obviously corresponds to the spatial infinity and it is irregular one of the rank $1$.

A solution of a differential equation near a regular singularity can be derived via the well-established Frobenius procedure. The Frobenius method gives rise to a series solution which is convergent in some domain near the singularity, but the convergence properties are out of the scope of this work. For an irregular singularity instead of the Frobenius method an asymptotic expansion can be applied.

To obtain the series solutions near the regular singularities an indicial equation should be written. The solution of the indicial equation, namely the indicial exponents determine the behaviour of solutions of the differential equation near corresponding singularity points. For the equation (\ref{rad_Ri_x}) it follows that near the point $x=0$ the indicial exponents are $\varkappa_1=0$ and $\varkappa_2=1/2-2\al_j$, for the point $x=1$ there are $\varkappa_3=0$ and $\varkappa_4=1/2-2\beta_j$. It is known that if the difference of indicial exponents for any regular singularity is noninteger then both solutions near the corresponding singularity is of power-law character. In the opposite case one of the solutions also contains a logarithmic contribution, but we do not focus on those subtle peculiarities.

Let us consider the domain around the point $x=0$ and take the first indicial exponent $\varkappa_1=0$, then we write the corresponding solution of the radial equation (\ref{rad_Ri_x}) as follows:
\begin{equation}\label{series_sol}
\tilde{{\cal R}}_{j}(x)=\sum^{+\infty}_{n=0}a^{(j)}_{n}x^{n}.
\end{equation}
Substituting the series (\ref{series_sol}) into the equation and after some transformations we obtain relations for the coefficients $a_n$ of the series. Namely, for the first two coefficients $a^{(j)}_0$ and $a^{(j)}_1$ we get:
\begin{equation}\label{a_1}
a^{(j)}_{1}=-\frac{\mu_j}{2\al_j+1/2}a^{(j)}_0.
\end{equation}
It is known that the coefficient $a^{(j)}_0$ can be chosen arbitrary, and we set it $a^{(j)}_0=1$ to equating our function to $1$ if $x=0$.
For the following coefficients $a^{(j)}_n$ we obtain a recurrent  relation of the form:
\begin{equation}\label{rec_coef}
a^{(j)}_{n+1}=\frac{1}{(n+1)(2\al_j+n+1/2)}\left[(n(2(\al_j+\be_j-\kappa)+n)-\mu_{j})a^{(j)}_{n}+(2\kappa(n-1)+\mu_{j}+\nu_{j})a^{(j)}_{n-1}\right].
\end{equation} 
The latter formula is in agreement with the relation for the $a^{(j)}_1$ coefficient, namely if we set $n=0$ and $a^{(j)}_{-1}=0$ it reduces to the relation (\ref{a_1}). The relation (\ref{rec_coef}) allows to derive explicit expressions for the coefficients $a^{(j)}_n$ of the Frobenius series (\ref{series_sol}), but we can also write a general solution of the equation (\ref{rad_Ri_x}) formally in terms of confluent Heun functions \cite{Fiziev_JPA10}:
\begin{multline}
\tilde{{\cal R}}_{j}(x)=C_{1}HeunC\left(2\kappa,2\al_j-\frac{1}{2},2\be_j-\frac{1}{2}, \delta_{j}, \eta_{j};x\right)+\\C_{2}x^{1/2-2\al_{j}}HeunC\left(2\kappa,\frac{1}{2}-2\al_j,2\be_j-\frac{1}{2}, \delta_{j}, \eta_{j};x\right),
\end{multline}
where $\delta_j=(r_{+}-r_{-})(A_{j}+B_{j})$, $\eta_j=\frac{3}{8}-(r_{+}-r_{-})B_{j}$ and here it is supposed that $1/2-2\al_j$ is noninteger. Similarly series solutions of the equation (\ref{rad_Ri_x}) which can be also expressed in terms of the Heun functions can be written around the other regular singularity $x=1$. 

Near the irregular singularity $x=\infty$ we use an asymptotic expansion for the radial wave-function $\tilde{{\cal R}}_j(x)$ of the form:
\begin{equation}\label{asymp_Rj}
\tilde{{\cal R}}_j(x)=e^{-\lambda_{j} x}x^{\rho_j}\sum_{n=0}^{+\infty}\frac{b^{(j)}_{n}}{x^n},
\end{equation}
where $\lambda_{j}$, $\rho_j$ and $b^{(j)}_{n}$ are unknown coefficients. To apply the asymptotic expansion (\ref{asymp_Rj}) and obtain the unknown coefficients it is necessary that factors near the wavefunction $\tilde{{\cal R}}_{j}(x)$ and its derivative $\tilde{{\cal R}}'_{j}(x)$ should be decomposed into asymptotic series as well, but the explicit form of the equation (\ref{rad_Ri_x}) shows that it can be done easily, namely we write:
\begin{gather}
2\kappa+\frac{2\al_{j}+1/2}{x}+\frac{2\be_{j}+1/2}{x-1}=2\kappa+\frac{2(\al_j+\be_j)+1}{x}+\sum^{+\infty}_{n=1}\frac{\left(2\be_{j}+1/2\right)}{x^{n+1}},\label{coef_1}\\
\frac{\mu_{j}}{x}+\frac{\nu_{j}}{x-1}=\frac{\mu_{j}+\nu_{j}}{x}+\sum^{+\infty}_{n=1}\frac{\nu_{j}}{x^{n+1}}.\label{coef_2}
\end{gather} 
Substituting the expansions (\ref{asymp_Rj}), (\ref{coef_1}) and (\ref{coef_2}) into the equation (\ref{rad_Ri_x}) we will be able to obtain the coefficient $\la_{j}$, $\rho_j$ and $b^{(j)}_n$. It is worth pointing out that the coefficient $b^{(j)}_0$ remains undefined and here there is a simailarity with $a^{(j)}_0$  coefficient in the Frobenius decomposition (\ref{series_sol}). For the parameters $\la_j$ and $\rho_j$ we find:
\begin{gather}
\la_{j}^2-2\kappa\la_{j}=0,\label{la_eq}\\
2(\kappa-\la_{j})\rho_{j}-(2(\al_{j}+\be_{j})+1)\la_{j}+\mu_{j}+\nu_{j}=0\label{rho_eq}.
\end{gather}
From the equation (\ref{la_eq}) we conclude that the parameter $\la_{j}$ takes two values $\la^{(1)}_{j}=0$ and $\la^{(2)}_j=2\kappa$, but remembering the relation ($\ref{ansatz_Ri}$), together with the asymptotic expansion (\ref{asymp_Rj}) we can conclude that the value $\la^{(2)}_{j}=2\kappa$ gives rise to the consequence that the radial part of the wave function changes its character, namely from ingoing to outgoing and vice versa and it can take place for instance for quasinormal modes, but it will not be studied here. Since we take only $\la^{(1)}_j=0$ we conclude that the asymptotic expansion (\ref{asymp_Rj}) can be chosen in a simpler form without the exponential factor.

Considering the equation (\ref{rho_eq}) and setting $\la_{j}=0$ we obtain an explicit relation for the exponent $\rho_j$, namely we write:
\begin{equation}
\rho_{j}=\frac{1}{2\kappa}(\mu_j+\nu_j)=\al_{j}+\be_{j}+\frac{1}{2}+\frac{1}{2\tilde{\kappa}}(A_j+B_j),
\end{equation}
where $\tilde{\kappa}=\pm i\omega$. It should be stressed that the exponent $\rho_{j}$ also depends on the type of the solution, namely there are different values for ingoing and outgoing particles. If we consider ingoing solution ($\tilde{\kappa}=i\omega$) and take $\al^{(1)}_j$, and $\be^{(1)}_j$ we obtain:
\begin{equation}
\rho^{(1)}_j=\frac{1}{2}\left(3+(-1)^{j+1}\right)+i\left((-1)^{j+1}-1\right)2m\omega(1+s^2_1+s^2_2).
\end{equation}
Now if $j=1$ the exponent $\rho^{(1)}_1=2$ and for $j=2$ the parameter $\rho^{(1)}_2$ becomes imaginary and it depends on $\omega$ as well as the black hole parameters. If now we take $\al^{(2)}_j$ and $\be^{(2)}_j$ we obtain:
\begin{equation}
\rho^{(2)}_j=\frac{1}{2}\left(1+(-1)^{j+1}\right)+i\left((-1)^j-1\right)2m\omega(1+s^2_1+s^2_2),
\end{equation}
and here the exponent $\rho^{(2)}_1$ is the same as $\rho^{(1)}_2$ and $\rho^{(2)}_2=1$. We also examine the outgoing solution ($\tilde{\kappa}=-i\om$) and in this case the exponent $\rho_j$ takes the following values:
\begin{eqnarray}
 \rho^{(1)}_j=\frac{1}{2}\left(3+(-1)^{j}\right)+i\left((-1)^{j+1}+1\right)2m\omega(1+s^2_1+s^2_2);\\
 \rho^{(2)}_j=\frac{1}{2}\left(1+(-1)^{j}\right)+i\left((-1)^j+1\right)2m\omega(1+s^2_1+s^2_2).
\end{eqnarray}
The analysis of the obtained relations gives rise to similar conclusions as for the ingoing solution.

Now using the asymptotic expansion (\ref{asymp_Rj}) we find the explicit relation for the coefficients $b^{(j)}_n$. In particular, for the coefficient $b^{(j)}_1$ we obtain:
\begin{equation}\label{b_1}
b^{(j)}_{1}=\frac{1}{\kappa}\left(\rho_{j}(\rho_{j}+2(\al_{j}+\be_{j}))+\nu_{j}\right)b^{(j)}_{0}.
\end{equation}
Finally, for the coefficient $b^{(j)}_{n-1}$ we write:
\begin{multline}\label{b_n_1}
b^{(j)}_{n-1}=\frac{1}{2(n-1)\kappa}\Big(((\rho_j-n+2)(\rho_j-n+2(\al_j+\be_j+1))+\nu_{j})b^{(j)}_{n-2}\\\left.+\sum^{n}_{l=3}\left((\rho_j-n+l)\left(2\be_{j}+\frac{1}{2}\right)+\nu_{j}\right)b^{(j)}_{n-l}\right).
\end{multline}
Here we point out that $n\geqslant 2$ and if $n=2$  the sum in the upper relation is not taken into account, namely in this case the relation (\ref{b_n_1}) reduces to the relation (\ref{b_1}) for $b^{(j)}_1$. 

It is known that for the standard Kerr metric separation of variables can be performed for wave equations for particles of arbitrary spin ($s=0,\pm\frac{1}{2},\pm1,\pm\frac{3}{2},\pm 2$). After the separation of variables the equations are usually written in the form of coupled Teukolsky master equations. As we have noted above the scalar field equation was considered in \cite{Cvetic_JHEP12}, but the wave equations for higher spin particles have not been examined yet. But taking into account the results for the scalar field \cite{Cvetic_JHEP12} and together with ours we can conjecture a generalization of the Teukolsky equations \cite{Teukolsky_PRL72,Teukolsky_ApJ73}, namely we write:
\begin{gather}
\label{mast_1}\frac{1}{\sin{\theta}}\frac{\partial}{\partial\theta}\left(\sin{\theta}\frac{\partial {\cal S}_{s}}{\partial \theta}\right)+\left(a^2\omega^2\cos^2{\theta}+2sa\omega\cos{\theta}-\frac{k^2+s^2+2ks\cos{\theta}}{\sin^2{\theta}}+E_{k,s}\right){\cal S}_{s}=0,\\\label{mast_2}
X^{-s}\frac{\partial}{\partial r}\left(X^{s+1}\frac{\partial {\cal R}_{s}}{\partial r}\right)+\left(\frac{K^2(r)-2is(r-m)K(r)}{X}+2is\omega F'(r)-\lambda_{k,s}\right){\cal R}_{s}=0,
\end{gather}
where $K(r)=F(r)+ka$, $\lambda_{k,s}=E_{k,s}+a^2\omega^2+2a\omega k-s(s+1)$ and supposedly the spin $s$ takes all the allowed values similarly as in the standard Teukolsky equation, namely $s=0,\pm\frac{1}{2},\pm1,\pm\frac{3}{2},\pm 2$. We point out that the angular master equation completely coincides with its counterpart for the Kerr background and this fact is clear, because even our angular system (\ref{ang_sys})
is identical to its Kerr counterpart. The radial master equation (\ref{mast_2}) is a slight generalization of the corresponding equation for the Kerr case and in the limit when the charges go to zero ($s_1=s_2=0$) it reduces to the radial Teukolsky equation. It should be also stressed that wave functions ${\cal S}_{s}(\theta)$ and ${\cal R}_{s}(r)$ in the equations (\ref{mast_1}) and (\ref{mast_2}) may not coincide exactly with the functions ${\cal S}_j(\theta)$ and ${\cal R}_j(r)$ directly, they are equal up to simple angular and radial dependent factors respectively.
\section{Remarks about wavefunction for the massive particle}
Due to indendent interest in the massive case here we consider radial equations for the massive particle. Angular part can also be examined, but it is completely identical to the standard Kerr case. We point out that in this section we rather describe main peculiarities of the massive radial equation, namely we compare it with the corresponding equation for the Kerr background, more detailed analysis will be examined elsewhere.   The decoupling of the radial system in the massive case gives rise to the following equations for the radial components ${\cal R}_j(r)$:
\begin{multline}\label{R1}
{\cal R}''_j+\left(\frac{X'}{2X}\mp\frac{i\mu_{e}\left(r_1+r_2)\right)}{2\chi\left(\lambda\pm i\mu_{e}\chi\right)}\right){\cal R}'_j+\frac{1}{X}\left.\Big(\pm i\omega F'+(\omega F+ka)\times\right.\\\left.\left(\frac{1}{X}\left(\omega F+ka\mp\frac{i}{2}X'\right)+\frac{\mu_e(r_1+r_2)}{\chi(\lambda\pm i\mu_e\chi)}\right)-\mu^2_e\chi^2-\lambda^2\right){\cal R}_j=0,
\end{multline}
where $j=1,2$ and similarly as above the upper and lower signs correspond to the function ${\cal R}_1$ and ${\cal R}_2$ respectively. In the Kerr limit ($s_1=s_2=0$) the later equation reduces to the standard radial equation on the corresponding background. For the dyonic black hole with equal charges $s_1=s_2$ the equation (\ref{R1}) also becomes simpler and its form is very close to the Kerr case.

We point out that for the mentioned above particular cases the equation (\ref{R1}) gains an additional regular singular point, which lies in the complex plane. For the nonequal charges we consider here ($s_1\neq s_2$) the situation is more subtle because the function $\chi$ is irrational. Since $\chi(r)$ is an irrational function, a specific coordinate transformation should be performed to rewrite the equation (\ref{R1}) in a canonical form. Namely, the following coordinate transformation can be used:
\begin{equation}\label{zr_transf}
\sqrt{(r+2ms^2_1)(r+2ms^2_2)}=mz-r.
\end{equation} 
From the latter relation we obtain:
\begin{equation}
r=\frac{m(z^2-4s^2_1s^2_2)}{2(z+s^2_1+s^2_2)}.
\end{equation}
The derivatives of the function ${\cal R}_{j}$ should be also rewritten, namely for the first derivative we write:
\begin{equation}\label{dR_dz}
\frac{\partial {\cal R}_j}{\partial r}=\frac{2(z+s^2_1+s^2_2)^2}{m(z+2s^2_1)(z+2s^2_2)}\frac{\partial {\cal R}_j}{\partial z},
\end{equation}
and for the second derivative we obtain:
\begin{multline}\label{d2R_dz2}
\frac{\partial^2 {\cal R}_j}{\partial r^2}=\frac{4(z+s^2_1+s^2_2)^4}{m^2(z+2s^2_1)^2(z+2s^2_2)^2}\frac{\partial^2 {\cal R}_j}{\partial z^2}+\\\frac{8(z+s^2_1+s^2_2)^3}{m^2(z+2s^2_1)^2(z+2s^2_2)^2}\left(1-\frac{(z+s^2_1+s^2_2)^2}{(z+2s^2_1)(z+2s^2_2)}\right)\frac{\partial {\cal R}_j}{\partial z}.
\end{multline}
Therefore we conclude that the transformation (\ref{zr_transf}) allows to rewrite the equation (\ref{R1}) in the form with rational coefficients only and then it can be reduced to a canonical form similarly as we did for the equation (\ref{rad_Ri_x}), but with additional singularities. We also note that apart of the technical side, which makes the corresponding transformations less tractable in comparison with the massless case, there is a conceptual point, because the transformation (\ref{zr_transf}) might bring additional the so-called apparent singular points for the radial equations (\ref{R1}), those apparent singularities should be carefully removed from further analysis, but it will be considered elsewhere. For the particular case $s_1=s_2$ or for the Kerr one the transformation (\ref{zr_transf}) becomes linear, which is a simple shift transformation. In the end, for the massive equation in the String frame (\ref{Dirac_SF}) the equation for the radial wave-function will be a bit simpler than (\ref{R1}), since as we have noted above in the radial system (\ref{rad_sys}) instead of the irrational function $\chi(r)$ we have $r_1=r+2ms^2_1$ and there is no need in a transformation like (\ref{zr_transf}), the equation will be similar to what we have for case if $s_1=s_2$. 
\section{Conclusions}
We have examined the four-dimensional Dirac equation on a rotating STU black hole background \cite{Cvetic_PRD96,Chong_NPB05} obtained in the framework of maximally supersymmetric Supergravity. Due to technical difficulties we consider only the pair-wise equal charges case. To recover the separabilty a specific contribution called torsion term is added to the standard Dirac equation, the source of the torsion is the Kalb-Ramond field, but further aspects of its properties remain to be examined. We should also note that the separability even of the massless Dirac equation in the background of the general four charges solution \cite{Cvetic_PRD96,Chong_NPB05} as well as for the eight charges black hole \cite{Chow_CQG14,Chow_PRD14_1,Chow_PRD14_2} remains an open problem. In comparison with the scalar field equations, we note that while the massless minimally coupled equation is separable for the general STU black hole background \cite{Cvetic_JHEP12_1,Cvetic_JHEP12}, the massive one is not. 

Even in the case of the pair-wise equal charges background geometry (\ref{BL_pair}), the Dirac equation reveals some specific features which do not appear for the Kerr background and they were not paid necessary attention to the Kerr-Sen solution \cite{Houri_JHEP10}. First, in comparison with the Kerr-Sen background this more general space-time requires that we have to consider corresponding generalization of the torsion form and the respectively modified Dirac equation becomes separable. The new peculiarity which has not been noted earlier, is the fact that the torsion form and corresponding modification of the Dirac equation are in general non unique giving rise to different physical systems. Namely, we have derived the torsion forms in two ways, the first one can be treated as ``geometrical'', it is used in the Einstein frame and it is based on a deformation or modification of Killing-Yano forms. The second way is more ``physical'', since it introduces the torsion as a field strength of the Kalb-Ramond two-form field potential which is an essential ingredient of String Theory. Taking into account quite general statements about conformal transformations of the generalized conformal Killing-Yano forms and the torsion forms \cite{Houri_JHEP10,Kubiznak_PLB09} we conclude that the torsion forms derived from different initial assumptions are not related to each other by a corresponding rescaling as it was initially expected.  Therefore, two approaches we have utilized to derive the torsion forms in general give rise to different modified Dirac equations. 

We want to point out that these two examples may be associated with different Dirac fields in the effective Supergravity of toroidally compactified String Theory \cite{Cremmer_PLB78,Cremmer_NPB79}. To have a fully-fledged explanation of this difference or to find out whether there is a relation between them it is necessary to examine fermionic sector of the effective Supergravity. But a very important conclusion can be derived even from the above studies, namely within the ``physical" approach where the torsion is associated with the Kalb-Ramond field or via corresponding dualization procedure it is related to the axion field, so even a neutral Dirac fermion conisidered as a probe allows to gain deeper insight of the Supergravity model in comparison with a minimally coupled scalar probe, since the scalar one is coupled just to the background geometry, whereas the fermion in addition has a coupling with the Kalb-Ramond or the axion fields depending on the preferable description.


The important result we would like to emphasize is the separability of the the massive Dirac equation in both frames, but it seems to be violated if the frame transformation is performed. The modified massless Dirac action shows its conformal invariance, similarly to the standard Dirac action if a corresponding rescaling of the Dirac spinors is performed. The conformal properties of the modified massless Dirac equation are considered explicitly for the types of torsion that we obtained, and proper rescaling of the Dirac spinors is shown. These results are also of crucial importance and as far as we know they have not been examined in earlier works.
 
We have also studied components of the Dirac wave-function. We extensively study the massless case, because of its relative simplicity and similarity to the Kerr case equation. It is shown that both radial and angular equations are reduced to the form of the confluent Heun equations, moreover for the angular part it is completely identical to the corresponding equation for the Kerr background. We have conjectured an analog of the Teukolsky equations that is true for $s=0$ minimally coupled scalar and $s=\frac{1}{2}$ Dirac fields, furhter analysis of the various forms of the conjectured generalization of the Teukolsky equation and some of its properties are studied in our forthcoming paper \cite{Cvetic_Liao}. Finally, we have made a brief sketch analyzing the massive radial equation, pointing out main peculiarities and leaving this technically more sophisticated problem for further study elsewhere. 
\section*{Acknowledgments}
{We thank Chris Pope,  Bernard Whiting and Haoyu Zhang for useful discussions. MMS is partially supported by the Fulbright Program grant for visiting scholars. MC is partially supported by the Slovenian Research Agency (ARRS No. P1-0306) and Fay R. and Eugene L. Langberg Endowed Chair funds, by DOE Award (HEP) DE-SC0013528, by a University Research Foundation Grant at the University of Pennsylvania and by the Simons Foundation Collaboration grant $\# 724069$.}

\appendix
\section{Spin-connection $1$-forms for conformally related frames and conformal invariance of a generalized Dirac action}\label{app_A}
For convenience we give explict relations for spin-connection $1$-forms for both frames we consider in this work.  The Cartan equations (\ref{Cartan_eq}) can be utilized to derive the the spin-connection $1$-forms. Namely, taking into account the explicit expression for the tetrad (\ref{tetrad_0}) we write the explicit relations for the spin-connection in the Einstein frame:

\begin{align}
{\omega^{0}}_{1} &=\left(\frac{\sqrt{X}}{\De^{1/4}_0}\right)'e^0+\frac{a}{2\De^{3/4}_0}F'\sin{\th}{e^3};\\ {\omega^{0}}_{2}&=-\frac{a}{\De^{3/4}_0}\cos{\th}\left(a\sin{\th}e^{0}-\sqrt{X}e^{3}\right);\\ {\omega^{0}}_{3}&=\frac{a}{\De^{3/4}_0}\left(\frac{F'}{2}\sin{\th}e^{1}-\sqrt{X}\cos{\th}e^{2}\right);\\ {\omega^{1}}_{2}&=\frac{1}{\De^{1/2}_0}\left(\left(\De^{1/4}_0\right)_{,\theta}e^{1}-\sqrt{X}\left(\De^{1/4}_0\right)'e^{2}\right);\\ {\omega^{1}}_{3}&=\frac{F'}{2\De^{3/4}_0}\left(a\sin{\th}e^{0}-\sqrt{X}e^{3}\right);\\{\omega^{2}}_{3}&=\frac{a\sqrt{X}}{\De^{3/4}_{0}}\cos{\th}e^{0}-\frac{1}{\sin{\th}}\left(\frac{\sin{\th}}{\De^{1/4}_0}\right)_{,\theta}e^{3}.
\end{align}
Here prime $'$ denotes derivatives with respect to the radial coordinate $r$, and $()_{,\theta}$ means derivative with respect to angular variable $\theta$. Now using the tetrad (\ref{tetrad_0}) we are able to write explicit forms for the space-time components of the given above spin-connection $1$-form.

If there is a metric conformally related to a known one, what actually takes place for the Einstein and String frames, one can use the Cartan structure equations  as well to derive spin-connection in the new frame. Both frames have affine connections which are in agreement with corresponding metrics, but since the frames are conformally related it is easy to obtain explicit relations for affine connections or/and spin-connections for these two frames. For convenience of the reader we give a brief sketch which establishes these ties.

It is known that the Levi-Civita connection on a pseudo-Riemannian manifold is introduced so as it preserves the metric (it is called a metric connection or compatible with the metric) and it is torsionless. According to a general theorem of Riemannian geometry the Levi-Civita connection is unique. In terms of the components of the metric tensor the metric compatibility looks as follows:
\begin{equation}\label{metr_comp1}
\nabla_{\mu}g_{\la\ka}=\partial_{\mu} g_{\la\ka}-\Gamma^{\si}_{\mu\la}g_{\si\ka}-\Gamma^{\si}_{\mu\ka}g_{\la\si}=0,
\end{equation}
where symbol $\nabla$ denotes the Levi-Civita connection or in other terms the covariant derivative, and $\Gamma^{\si}_{\rho\tau}$ is used to denote affine connection coefficients (symbols). Now imposing the torsionless condition ($\Gamma^{\si}_{\rho\tau}=\Gamma^{\si}_{\tau\rho}$) and using the relations (\ref{metr_comp1}) we derive the explicit expressions for the coefficients $\Gamma^{\si}_{\rho\tau}$ called the Christoffel symbols. 

Similarly, the torsionless metric connection $\tilde{\nabla}$ is introduced for the conformally related metric $\tilde{g}_{\mu\nu}$, namely we write:
\begin{equation}\label{metr_comp2}
\Tilde{\nabla}_{\mu}\tilde{g}_{\la\ka}=\partial_{\mu}\tilde{g}_{\la\ka}-\tilde{\Gamma}^{\si}_{\mu\la}\tilde{g}_{\si\ka}-\tilde{\Gamma}^{\si}_{\mu\ka}\tilde{g}_{\la\si}=0.
\end{equation}
To make our analysis quite general here we do not focus on a particular form of the conformal transformation (\ref{conf_rel}) together with the factor (\ref{dilat}) related to the problem we consider, but we assume that the relations between the metric is of the form:
\begin{equation}\label{conf_2}
\Tilde{g}_{\mu\nu}=e^{2\Phi}g_{\mu\nu},
\end{equation}
where the scaling factor $\Phi=\Phi(x^{\ka})$ is a function of space-time coordinates. Taking into account the relations (\ref{metr_comp1})-(\ref{conf_2}) we obtain relations between the Christoffel symbols for two conformally related frames, namely we write:
\begin{equation}\label{chr_symb}
\tilde{\Gamma}^{\ka}_{\mu\nu}=\Gamma^{\ka}_{\mu\nu}+\Phi_{,\mu}\delta^{\ka}_{\nu}+\Phi_{,\nu}\delta^{\ka}_{\mu}-\Phi_{,\si}g^{\si\ka}g_{\mu\nu},
\end{equation}
where $\Phi_{,\mu}$ is the partial derivative over $x^{\mu}$.  To establish a relation for the spin-connections for conformally related frames we use the frame compatibility condition, which in terms of components looks as follows:
\begin{equation}\label{comp_frame}
\nabla_{\mu}e^{A}_{\nu}=\partial_{\mu}e^{A}_{\mu}-\Gamma^{\si}_{\mu\nu}e^{A}_{\si}+{\omega^{A}_{\mu B}}e^{B}_{\nu}=0, 
\end{equation}
and here $e^{A}_{\mu}$ are frame fields (tetrad) components ($e^{A}=e^{A}_{\mu}dx^{\mu}$) and ${\omega^{A}_{\mu B}}$ are the components of the spin-connection $1$-form (${\omega^{A}}_{B}={\omega^{A}_{\mu B}}dx^{\mu}$).
For the conformally related frame the compatibility condition can be written in a similar way:
\begin{equation}\label{comp_frame2}
\tilde{\nabla}_{\mu}\tilde{e}^{A}_{\nu}=\partial_{\mu}\tilde{e}^{A}_{\mu}-\tilde{\Gamma}^{\si}_{\mu\nu}\tilde{e}^{A}_{\si}+{\tilde{\omega}^{A}_{\mu B}}\tilde{e}^{B}_{\nu}=0. 
\end{equation}
The relation (\ref{conf_2}) implies a simple relation between the frame fields (tetrads), namely:
\begin{equation}\label{frame_rel}
\tilde{e}^{A}=e^{\Phi}e^{A} \quad \Leftrightarrow \quad \tilde{e}^{A}_{\mu}=e^{\Phi}e^{A}_{\mu}.
\end{equation} 
Taking into account the relations (\ref{chr_symb})-(\ref{frame_rel}) and after simple transformations we obtain the relation for the spin-connection components for two conformally related frames, namely we write:
\begin{equation}\label{spin_conn}
\tilde{\omega}^{A}_{\mu B}={\omega}^{A}_{\mu B}+\Phi_{,\si}\left(e^{\si}_{B}e^{A}_{\mu}-e^{\si A}e_{\mu B}\right).
\end{equation}
We point out here that the upper Lorentzian index in the latter relation can be dropped down and the resulting relation manifests  the antisymmetry of the spin-connection coefficients with respect to the Lorentzian indices.

Finally we write the relations for the spin-connection $1$-forms in the String frame which, as we pointed out above can be derived directly from the Cartan structure equations or using the relations (\ref{spin_conn}):  
\begin{align}
{\tilde{\omega}^{0}}_{1}&=\frac{\De^{1/2}_{0}}{R^2_1}\left(\frac{R_1\sqrt{X}}{\De^{1/2}_{0}}\right)'\tilde{e}^{0}+\frac{aF'\sin{\theta}}{2R_1\De^{1/2}_{0}}\tilde{e}^3;\\
{\tilde{\omega}^{0}}_{2}&=\frac{a\cos{\theta}}{R_1}\left(-\frac{a\sin{\theta}}{R^2_1}\tilde{e}^{0}+\frac{\sqrt{X}}{\De^{1/2}_{0}}\tilde{e}^3\right);\\ {\tilde{\omega}^{0}}_{3}&=\frac{a}{R_1\De^{1/2}_{0}}\left(\frac{1}{2}F'\sin{\theta}\tilde{e}^{1}-\sqrt{X}\cos{\theta}\tilde{e}^2\right);\\ {\tilde{\omega}^{1}}_{2}&=\frac{1}{R^2_1}\left(\left(R_1\right)_{,\theta}\tilde{e}^1-\sqrt{X}R'_{1}\tilde{e}^2\right);\\ {\tilde{\omega}^{1}}_{3}&=\frac{1}{R_1}\left(\frac{aF'\sin{\theta}}{2\De^{1/2}_{0}}\tilde{e}^{0}-\frac{\sqrt{X}R'_{1}}{R_1}\tilde{e}^{3}\right);\\{\tilde{\omega}^{2}}_{3}&=\frac{1}{R_1}\left(\frac{a\sqrt{X}}{\De^{1/2}_0}\cos{\theta}\tilde{e}^{0}-\frac{\De^{1/2}_{0}}{R_1\sin{\theta}}\left(\frac{R_1\sin{\theta}}{\De^{1/2}_0}\right)_{,\theta}\tilde{e}^{3}\right).
\end{align}

It is known that the standard massless Dirac action is invariant under a conformal rescaling of the metric if the Dirac spinors are properly rescaled. Here, using the transformation relation for the torsion forms  \cite{Houri_JHEP10} we show that the torsion modified Dirac action is also invariant under the conformal rescaling.

To have our discussion self-contained we firstly show conformal invariance of the standard Dirac action, even though this fact is well-known. Let us write the standard Dirac action (for any dimension $n$), and consider the massless case:
\begin{equation}\label{Dirac_act}
S_{D}=\int d^{n}x\sqrt{-g}\bar{\Psi}\hat{\gamma}^{\mu}\left(\partial_{\mu}+\Gamma_{\mu}\right)\Psi,
\end{equation}
Here $\Psi$ and $\bar{\Psi}$ are the Dirac spinor and its adjoint, the space-time gamma matrices $\hat{\gamma}^{\mu}$ and the spinor connection $\Gamma_{\mu}$ are defined by the relations (\ref{gamma_matr}) and (\ref{spinor_con}) respectively.
We rewrite the action in the conformally related frame, defined by the relations (\ref{conf_2}) for the metrics or by (\ref{frame_rel}) for the frame fields. Using the relation (\ref{spin_conn}) we can write:
\begin{equation}
\omega_{\mu AB}=\tilde{\omega}_{\mu AB}-\Phi_{,\sigma}\left(\tilde{e}^{\sigma}_{B}\tilde{e}_{\mu A}-\tilde{e}^{\sigma}_{A}\tilde{e}_{\mu B}\right).
\end{equation}
Contracting the connection components $\omega_{\mu AB}$ with the Lorentzian and the space-time gamma matrices we obtain:
\begin{equation}\label{contr_tr}
\hat{\gamma}^{\mu}\Gamma_{\mu}=e^{\Phi}\hat{\tilde{\gamma}}^{\mu}\left(\tilde{\Gamma}_{\mu}-\frac{n-1}{2}\Phi_{,\mu}\right),
\end{equation}
where $\hat{\tilde{\gamma}}^{\mu}=\tilde{e}^{\mu}_{A}\gamma^{A}=e^{-\Phi}\hat{\gamma}^{\mu}$ and $\tilde{\Gamma}_{\mu}=\frac{1}{4}\tilde{\omega}_{\mu AB}\hat{\gamma}^{A}\hat{\gamma}^{B}$ are the curved gamma matrices and spinor connection coefficients for the conformally related metric $\tilde{g}_{\mu\nu}$. Substituting the upper relation into the Dirac action (\ref{Dirac_act}) and rewriting the determinant of the metric tensor we obtain:
\begin{equation}
S_D=\int d^{n}x\sqrt{-\tilde{g}}e^{(1-n)\Phi}\bar{\Psi}\hat{\tilde{\gamma}}^{\mu}\left(\partial_{\mu}+\tilde{\Gamma}_{\mu}-\frac{n-1}{2}\Phi_{,\mu}\right)\Psi.
\end{equation}
If the spinors are rescaled as follows: 
\begin{equation}\label{spinor_resc}
\Psi=e^{(n-1)\Phi/2}\tilde{\Psi},\quad \bar{\Psi}=e^{(n-1)\Phi/2}\bar{\tilde{\Psi}}\quad \Leftrightarrow \quad \tilde{\Psi}=e^{-(n-1)\Phi/2}\Psi,\quad \bar{\tilde{\Psi}}=e^{-(n-1)\Phi/2}\bar{\Psi}.
\end{equation}
Then, the latter action can be rewritten in the form:
\begin{equation}
S_D=\int d^{n}x\sqrt{-\tilde{g}}\bar{\tilde{\Psi}}\hat{\tilde{\gamma}}^{\mu}\left(\partial_{\mu}+\tilde{\Gamma}_{\mu}\right)\tilde{\Psi}.
\end{equation}
Therefore, we demonstrated the conformal invariance of the standard Dirac action.

To derive the generalized Dirac equation modified by the torsion (\ref{D_eq_mod}), we assume that corresponding generalized Dirac action takes the form:
\begin{equation}\label{GD_action}
S_{GD}=\int d^{n}x\sqrt{-g}\bar{\Psi}\hat{\gamma}^{\mu}\left(\partial_{\mu}+\Gamma_{\mu}-\alpha T_{\mu\nu\lambda}\hat{\gamma}^{\nu}\hat{\gamma}^{\la}\right)\Psi,
\end{equation}
where $T_{\mu\nu\la}$ are the space-time components of the torsion form and the coefficient $\alpha$ is introduced instead of the combinatorial factor $1/24$. Since we have already shown the conformal invariance of the standard Dirac action, it is necessary to show the conformal invariance of the torsion-related contribution. It was proven in \cite{Houri_JHEP10} that the torsion form $T_{\mu\nu\la}$ transforms under conformal rescaling of the metric (\ref{conf_2}) as follows  $\tilde{T}_{\mu\nu\la}=e^{2\Phi}T_{\mu\nu\la}$, taking into account transformations of the spinors (\ref{spinor_resc}), the gamma matrices and the determinant of the metric we obtain:
\begin{equation}
\int d^{n}x\sqrt{-g}\bar{\Psi}\left(-\alpha T_{\mu\nu\la}\hat{\gamma}^{\mu}\hat{\gamma}^{\nu}\hat{\gamma}^{\la}\right)\Psi=\int d^{n}x\sqrt{-\tilde{g}}\bar{\Psi}\left(-\alpha \tilde{T}_{\mu\nu\la}\hat{\tilde{\gamma}}^{\mu}\hat{\tilde{\gamma}}^{\nu}\hat{\tilde{\gamma}}^{\la}\right)\tilde{\Psi}.
\end{equation}
We see that the torsion-related contribution into the action (\ref{GD_action}) is conformally invariant, therefore this action is conformally invariant as well.

We point out that to ensure conformal invariance of the Dirac action for a massive fermion, the fermion mass should be rescaled as $\mu=e^{\Phi}\tilde{\mu}$, therefore if for one of the frames the mass is a constant in the conformally related metric it turns to be a function of coordinates. This fact emphasizes one of important features of conformal rescaling, namely if instead of the constant fermion mass $\mu_e$ in a particular frame there is a function of coordinates it is possible to find a convenient  frame conformally related to the previous one which allows if not transform this function into a constant, but obtain at least a simpler function.   
 
\section{Dualization of the axion field and the Legendre transformation of the truncated bosonic Lagrangian}\label{app_B}
Higher form fields, in particular Kalb-Ramond fields play important role in String Theory and Supergravity. They are essential parts for  supersymmetry transformations of fermionic fileds \cite{Nishino_PLB84, Salam_PLB84} and consequently there is nontrivial coupling between the Kalb-Ramond and  the fermion fields reflected in the corresponding equations of motion. They are also important in order to define the Killing spinor equations for String Theory or Supergravity backgrounds \cite{Frolov_LivRevRel17,Chow_CQG17}. A decade ago it was shown that they are necessary in order to obtain the separable Dirac equations on the Kerr-Sen space-time \cite{Houri_JHEP10}. 

Due to its importance for our problem we show  that Kalb-Ramond fields can be introduced via a Legendre transformation of the Lagrangian (\ref{Lagr_1}). In general this procedure is well-established, in particular in four dimensional case it is given in terms of components in the seminal paper \cite{Sen_IJMPA94}. But to make our study self-contained we describe it in terms of the differential forms, it makes the connection between the axion field and Kalb-Ramond fields more transparent.

The Kalb-Ramond field can be defined as a combination of an exterior derivative of a $2$-form $B$-field (potential) and ``Chern-Simons''-like terms. Due to the fact that in our setup there are two independent gauge fields defining electric and magnetic components respectively we define the Kalb-Ramond field as follows:
\begin{equation}\label{H_def}
H=dB-\frac{1}{2}\left(A_1\wedge F_1+A_2\wedge F_2\right).
\end{equation}  
Taking its exterior derivative and using the fact that $d^2B=0$ and $dF_{i}=0, i=1,2$  we write:
\begin{equation}\label{dH_der}
dH=-\frac{1}{2}\left(F_1\wedge F_1+F_2\wedge F_2\right).
\end{equation}
We see that this is the very same expression(up to a multiplication factor $\chi$) that shows axion-gauge fields coupling in the Lagrangian (\ref{Lagr_1}). Thus, the Lagrangian (\ref{Lagr_1}) can be rewritten in the following form:
\begin{equation}\label{Lagr_2}
{\cal L}=R*1-\frac{1}{2}*d\phi\wedge d\phi-\frac{1}{2}e^{2\phi}*d\chi\wedge d\chi-\frac{1}{2}e^{-\phi}\left(*F_1\wedge F_1+*F_2\wedge F_2\right)+\chi dH.
\end{equation}
The exterior derivative in the last term of the upper Lagrangian can be moved to $\chi$ as follows:
\begin{equation}\label{Legendre_tr}
\chi dH=d(\chi H)-d\chi\wedge H,
\end{equation}
and since the total exterior derivative can be omitted the Lagrangian (\ref{Lagr_2}) can be represented in the form:
\begin{equation}\label{Lagr_3}
{\cal L}=R*1-\frac{1}{2}*d\phi\wedge d\phi-\frac{1}{2}\left(e^{2\phi}*d\chi-2H\right)\wedge d\chi-\frac{1}{2}e^{-\phi}\left(*F_1\wedge F_1+*F_2\wedge F_2\right).
\end{equation}
Since the $B$-field in the definition (\ref{H_def}) was not explicitly given, there is still ``freedom'' in the definition of the Kalb-Ramond field. Now we assume that:
\begin{equation}\label{H_chi}
 H=e^{2\phi}*d\chi \quad \Leftrightarrow \quad d\chi=e^{-2\phi}*H.
\end{equation}
Using the upper relation we finally rewrite the Lagrangian (\ref{Lagr_3}) in the form with the explicit Kalb-Ramond term:
\begin{equation}\label{Lagr_H}
{\cal L}=R*1-\frac{1}{2}*d\phi\wedge d\phi-\frac{1}{2}e^{-2\phi}*H\wedge H-\frac{1}{2}e^{-\phi}\left(*F_1\wedge F_1+*F_2\wedge F_2\right).
\end{equation}
The obtained form of the Lagrangian (\ref{Lagr_H}) is in perfect agreement with its component representation given in earlier papers \cite{Cvetic_NPB96_1,Cvetic_PRD96,Cvetic_NPB96,Sen_IJMPA94}. We also note that in contrast with the axion field $\chi$ the Kalb-Ramond field $H$ in (\ref{Lagr_H}) is not supposed to be a fundamental one (in the sense that we assume it is basic to derive the equations of motion), the fundamental field here is the field $B$ which is not given explicitly, but it can be obtained using the relations (\ref{H_def}) and (\ref{H_chi}). The  crucial thing we would like to stress is the fact that the transformation of the Lagrangian we have made is nothing else but a Legendre transformation, namely the same transformation is given by (\ref{Legendre_tr}) and  both ``initial'' (\ref{Lagr_1}) and ``final'' (\ref{Lagr_H}) Lagrangians give rise to completely equivalent descriptions of the system as it has to be. 

Finally, rather technical, but a very important detail, if a Legendre transformation for a particular value has been done then the equations of motion for the transformed value are satisfied as identities, it is nothing else but the consistency check. Namely, in our case, the equation of motion for the axion field $\chi$ derived from the Lagrangian (\ref{Lagr_1}) is as follows:
\begin{equation}
d\left(e^{2\phi}*d\chi\right)+\frac{1}{2}\left(F_1\wedge F_1+F_2\wedge F_2\right)=0.
\end{equation}
Now, if one uses the relations (\ref{H_chi}) and (\ref{dH_der}) one can see that the upper equation becomes an identity.

Using the relation (\ref{H_chi}) we calculate the Kalb-Ramond field firstly in the Einstein frame:
\begin{equation}\label{H_EF}
H=\frac{a(r_2-r_1)}{\De^{5/4}_{0}}\left((r^2_1-a^2\cos^2{\th})\sin{\th}e^0\wedge e^1\wedge e^3-2r_1\sqrt{X}\cos{\th}e^0\wedge e^2\wedge e^3\right).
\end{equation}
In the String frame it can be written in the form:
\begin{equation}\label{H_SF}
H=\frac{a(r_2-r_1)}{R^3_1\De^{1/2}_{0}}\left((r^2_1-a^2\cos^2{\th})\sin{\th}\tilde{e}^0\wedge \tilde{e}^1\wedge \tilde{e}^3-2r_1\sqrt{X}\cos{\th}\tilde{e}^0\wedge \tilde{e}^2\wedge \tilde{e}^3\right).
\end{equation}
The latter relation can be recast as follows:
\begin{equation}
H=-\frac{a\sin{\th}}{R_1}\left(\frac{r_1+r_2}{\De^{1/2}_{0}}-\frac{2r_1}{R^2_1}\right)\tilde{e}^0\wedge \tilde{e}^1\wedge \tilde{e}^3+\frac{2a\sqrt{X}\cos{\th}}{R_1}\left(\frac{1}{\De^{1/2}_0}-\frac{1}{R^2_1}\right)\tilde{e}^0\wedge \tilde{e}^2\wedge \tilde{e}^3.
\end{equation}
It is easy to see that the written above relation up to a sign coincides with the torsion $\tilde{T}_{-}$ (\ref{torsion_SF}), the difference in sign, as we have noted above, is caused by the fact that the Killing-Yano tensors are defined up to the overall sign.

\end{document}